 \documentclass[prd,a4paper,
twocolumn,amsmath, amssymb, floatfix,
nofootinbib,wrapfloat byrevtex]{revtex4}

 \usepackage{amsmath,mathrsfs,upgreek,graphicx,placeins,float}
 \usepackage{color}
 \usepackage{booktabs}
 \usepackage{multirow}
 
 \numberwithin{equation}{section}

 \newcommand{\beq}[1]{\begin{equation}\label{#1}}
 \newcommand{\eeq}{\end{equation}}
 \newcommand{\bea}[1]{\begin{eqnarray}\label{#1}}
 \newcommand{\eea}{\end{eqnarray}}
  \newcommand{\bec}[1]{\begin{center}\label{#1}}
 \newcommand{\eec}{\end{center}} 
 \newcommand{\mm}{{\rule[1.5pt]{3pt}{0.6pt}}}

\begin{document}

 \title{Microscopic State of BHs and an Exact One Body Method for Binary Dynamics in General Relativity}
 \author{Ding-fang Zeng}
 \email{dfzeng@bjut.edu.cn}
 \affiliation{Beijing University of Technology, Chaoyang, Bejing 100124, P.R. China,
 \\
 Niels Bohr Institute, Blegdamsvej 17, 2100 Copenhagen, Denmark}

 \begin{abstract} 
In gravitational collapses, the horizon and singularity's realisation in the finite future of the proper time used co-moving observer happens in the future of infinitely far away future of the normal time used outside probe. To the latter the horizon and singularity of a black hole formed through gravitational collapse are physical realities only in the sense of uncertainty principle and ensemble interpretation. We provide two exact time dependent solution families to the Einstein equation and show that they form a pair of complementarity description for the microscopic state of black holes by showing that the Bekenstein-Hawking entropy formula follows properly from their canonical wave function's degeneracy. We also develop an eXact One Body method for general relativity two-body dynamics whose conservative part requires no perturbative input from post newtonian approximation and applies to the full three stages of black hole binary merger events. By this method, we analytically calculate the gravitational wave forms following from such merger processes. In the case black holes carry exact and apriori horizon and singularity our wave forms agree with those following from conventional effective one body method but exhibit more consistent late time behaviour. In the case the black holes carry only asymptotic horizon and extended inner structure thus experiencing banana shape deformation as the merger progresses, our wave forms exhibit all features especially the late time quasi-normal mode type oscillation seen in real observations.
\end{abstract}
 \maketitle
 
\tableofcontents
 
\section{Introduction}

The existence of Bekenstein-Hawking entropy \cite{hawking1971,Bekenstein1972,Bekenstein1974,hawking1974,Jacobson1996,Wald1999} is a strong hint that black holes are inner-structured \cite{Bekenstein1974a,Mukhanov1986,Bekenstein1995,Bekenstein1997a,Bekenstein1997b,dfzeng2017,dfzeng2018a,dfzeng2018b,dfzeng2020,dfzeng2021,dfzeng2022,Allen1987,NambuSasaki1988,Nagai1989,BFZ1994,KMY2013,Stojkovic2008b,stojkovic2020unp}. The current observational data  \cite{GWTC1,GWTC2,GWTC3} on gravitational waves arising from Black Hole Binary (BHB) merger events may brought us information about such inner-structures already. The purpose of this work is to provide a thorough understanding of such inner-structures in the framework of standard general relativity and canonical quantum mechanics. At the same time an exact one body method will be developed to calculate the gravitational wave form of BHB merger process so that the effects of their inner-structures on observational signals can be exhibited directly. Both these two points seem to be impossible purposes. We will show that the first impossible arises from our wrong taking of the horizon and singularity as physical realities independent of probe and physical interactions. The second impossible arises from our wrong choosing of the effective background geometry which is aimed to account for the relative motion between the two bodies as static. This choice not only makes exact solutions to the problem impossible, but also the known effective one body method inconsistent at the principle level.

By physical reality, we refer to measurement outputs of any single type of physical quantity. In contrasts, we will use physical law to refer to relations bridging at least two types of physical realities. Physical reality is invariant under general coordinate transformation but observer dependent. Physical law is not only general coordinate invariant but also observer independent. Taking Einstein Equation $R_{\mu\nu}-\frac{1}{2}g_{\mu\nu}R=8\pi GT_{\mu\nu}$ as an example, both sides of it are physical realities.  We will give concrete examples to show how the physical law's observer independence is confused with the physical reality's observer dependence thus leading to the wrong belief of horizon and singularity's probe independence in section \ref{secPhysReality}. But to understand the analysis there firmly we need five sections to build the right description for the microscopic state of black holes. In a sense, all sections of this work are original and innovative. However, for those who only want to know why our two purposes are possible, they can jump to section \ref{secPhysReality}  and \ref{secXOBmethod} directly. We also have a short version of this long paper which is available on email requsting. The following is a highlight of the key point of each sections of this work.

Due to Penrose and Hawking's singularity theorem \cite{Penrose1965,Hawking1976,geroch1979,SingularityHawking}, inner-structure of black holes is very seldomly believed to have relevance with exact solutions to the four dimensional Einstein equation. Most of the string theory \cite{strominger1996,DasMathur9601,CallanMaldacena1996,HorowitzStrominger9602,MaldacenaSusskind9604,fuzzballsMathur2002,LuninMathur0202,fuzzballSkenderis2007,LuninMaldacenaMaoz0212,MT0103,GMathur0412,mathur0502,mathur2005,Jejjala0504,KST0704,Mathur0706} and loop quantum gravity researchers \cite{lqgEntropy1996,lqgEntropy1997,AshtekarBojowald0509,Modesto0509,Campiglia0703,Modesto0811,Modesto0611,Corichi1506,Bohmer0709,rovelli2015,MMP2021} believe that in the framework of general relativity, as long as falling into the horizon, the fate of all matters is hitting on the singularity and becoming a part of it. So the only existence form of matters inside the horizon is a singular point characterised by mass, charge and spin parameters exclusively.  So the most general microscopic structure allowed by general relativity is those exhibited in the Kerr-Neumann metric.

However, just as any existence question in philosophies do, the form of matter's existence in physics is also related with the definition of time. In general relativity, this is especially so. Because of time delay effects caused by gravitation, the outside fixed position observers cannot see the horizon's formation in a to be black hole collapsing star in any finite future quantifiable with their physical time definition. These observers are not required to be human being or any bio-wisdom. They can be any outside probe or detectors with fixed position or approximately fixed position. For example, either participant involved in a merger event of BHB system works as probes of its inspiral partner. These probes will not see the horizon's formation in their partners' evolution in any finite future quantifiable with their physical time definition. So to them, the form of matter's existence inside their merger partners is just that of the frozen star \cite{frozenStar1971,Buchdahl1959,chand1964a,chand1964b,bondi1964}. But the reason for being frozen here is the infinite gravitational time delay due to the horizon to form, instead of any exclusive force such as that arising from the neutron degeneration pressure or any other unknown physics \cite{BrusteinFrozen2301,Brustein2304,Brustein2310}. We will provide exact metric and Penrose-Carter diagram description for this frozen star in section \ref{secMetricOFPO} of this work.

By the freely falling observer's time definition, materials consisting of a to be black hole collapsing star would indeed contract into the mathematical surface defined by $r=2GM$ in the finite future. However, both the materials being observed and the observer themselves experience not any exotics as that surface is being walked through, all of them are freely falling. As {\bfseries\em local} observer and observable, the spacetime region around them are completely flatten. The non-zero curvature is physically relevant only to the global viewpoint holden investigators. This is also the case even when the $r=0$ point is concerned. The observer and observables coming from the east world of the central point just normally walk over that point and get to the west world afterwards. Tidal forces affect only {\bfseries\em non-local} or finite sized observer and observables. However, general relativity is a time-reverse invariant theory. No matter how strongly is an extended object tidally stretched as it walking to the $r=0$ point, its shape and inner structure will be retrieved when it walking away from that point just as the world is time reversed \cite{Turok1510,Turok1612,CPTsymmUniverse2018,CPTsymmUniverse2022,Turok2208}. Penrose and Hawking's singularity theorem \cite{Penrose1965,Hawking1976,geroch1979,SingularityHawking} only requires that all matter contents get into the horizon onto the singularity (an equal time hyper surface) in finite proper time. It does not prohibit them from walking across that epoch. Exact metric description and PC-diagram representation for this picture will be provided in section \ref{secMetricComObserver}. 

With section \ref{exampleQuantisation} supplemented as a working example, section \ref{secQuantisation}  discusses the quantisation of inner-structures described by the exact metric families of sections \ref{secMetricOFPO} and \ref{secMetricComObserver}. We will decompose the matter content of spherically symmetric black holes into many concentric shells characterised by each shell's mass and radial quantum number and write the wave functional of the whole system as direct product of all compositing shells. By considering all possible shell decomposition schemes allowed by the total mass summation rule and the radial quantum number assigning schemes allowed by the asymptotic horizon condition, we analytically show that the degeneracy of the system's wave functional is exactly that required by Bekenstein Hawking entropy formula, except for some logarithmic type correction. Numerically determining this correction is possible but challenging as the mass of the black hole becomes large.  
In contrast with the euclidean path integration method based on saddle point approximation \cite{entropyAshokeLogCorrec2013,GibbonsHawking1977,BrownYork1994} and string theory calculation \cite{Ashoke0504,Ashoke0506,Ashoke0508,Ashoke0611,Ashoke0708,Ashoke1005,Ashoke1008,Ashoke1106,Ashoke1108,Ashoke1109,Ashoke1204,Ashoke1402,Ashoke1405,Ashoke2306}, our microscopic state definition and number counting for black holes are purely general relativity based and resort not any hyper physical concepts such as supersymmetry or extra spatial dimension. But our black holes will exhibit similar observational features as string theory fuzz balls \cite{mayerson2021fuzzball,mayerson2022microReview} if their shadow images are taken \cite{joao2302,GLM2008} or gravitational wave echos \cite{gwEchoCardoso1902,gwEchoMaggio1907,gwEchoKaloper1912,gwEchoCardoso2007,gwEchoStojkovic2011,gwEchoMaggio2202} are measured.

The black hole complementarity principle \cite{complementarity1990thooft,complementarity1993verlinde,complementarity1993schoutens,complementarity1993,complementarity1994} believes that both the outside fixed position observer and the freely falling observer's description of a collapsing star's evolution are equally valid. We will elaborate in section \ref{secComplementarity} that, for the outside fixed position observers, the uncertainty principle or quantum fluctuation's ubiquity implies that parallel universe or ensemble description of the initial status of materials consisting of the black hole is necessary. While for the freely falling co-moving observers, this necessity can be interpreted as the ergodicity of the system's evolution on the proper time axis. So when the initial status' uncertainty or the evolution dynamics' ergodicity is considered, not only are the two types of observer's description both valid, but also are they complete. The fire wall paradoxes \cite{fireworksAMPS2012,fireworksAMPS2013,EREPR} will be analysed in this section to show that the truly rational resolution to the information missing puzzle \cite{fuzzball2009mathur,fuzzball2021mathur,NVW2013a,NVW2013b,giddings2022,Traschen,Stojkovic2007,Stojkovic2008a,Stojkovic2013,Stojkovic2014,Stojkovic2015} is to give up the popular vacuum fluctuation and partial escaping mechanism for Hawking radiation instead of the idea of black hole complementarity. This section mainly consists of logic reasoning instead of mathematical derivation. So we put it in the middle part of the paper to avoid preaching. Mathematical derivations underlying can be found in our earlier works \cite{dfzeng2021,dfzeng2022}.

Through three concrete examples, section \ref{secPhysReality} will analyse how the belief that gravitational collapse will cause the formation of observer independent horizon and singularity makes the wrong of (A) taking the horizon and singularity of global viewpoint holders as the horizon and singularity detectible to local viewpoint holders and (B) taking the physics law's general coordinate invariance as the physical reality's observer or probe independence. The physical reality's observer or probe independence is not the fact at all. Our three examples are, (i) the frequency of light signals emitted from the surface of a collapsing star and measured by an outside fixed position observer and a freely falling observer; (ii) the pressure of a dust type test mass shell freely falling towards the horizon of a pre-existing Schwarzschild black hole measured by a far away fixed position observer and a freely falling observer co-moving with the shell itself and (iii) the horizon of two cosmological sized super giants separated double hubble radius away in our real universe. This section is the key part of this work but we burry it here because it is conceptually so bold that some people may think it non-sense at all.

Section \ref{secBanana} will show that when the horizon is taken as a probe dependent physical reality, the black holes involved in the binary merger event will experience banana shape deformation under the inhomogeneous back reaction force of gravitational wave radiation. We will illustrate such a deformation's occurrence in the newtonian gravitation theory and argue that the physical picture would be similar in the fully general relativistic treatments of the system's evolution. We then provide a static mechanical analysis for the banana shape deformation of black holes and show that the radiation activity of such BHBs will be suppressed by a factor of $\frac{\sin4GMz/a(t)}{4GMz/a(t)}$ when they inspiral and merge. Here $z$ is the system's stretching mulriple relative to the diameter of the undeformed Schwarzschild horizon. $z$ is time dependent whose concrete form reflects the black hole's deformation sensitivity or susceptibility under the back reaction force of gravitational wave radiation.

Section \ref{secXOBmethod} will propose an exact one body (XOB) method to calculate the relative motion orbit of general relativity two-body systems. This method will avoid the self-contradicting ingredients of conventional effective one body (EOB) method \cite{eob1998,eob2000,eob2009}. The conservative part of XOB hamiltonian involves no post-newtonian approximation \cite{Blanchet9501,Damour0005,PatiWill0007,blanchet0209,pnApproxLivingReview} as inputs. Taking the most simple quadrupole radiation of gravitational waves as the source of dissipation, this method will allow us to trace the relative motion of binary systems even when their evolution develops to the ring-down stage. By this method we will calculate the relative motion orbit and gravitational wave forms for binary systems consisting of standard Schwarzschild black holes to make comparisons with the results of conventional EOB method. Our results will coincide with that of EOB at early times but behaves more rationally as the very late time stage is arrived. By looking the black hole as an arc of fixed length, this method will yield gravitational wave forms highly similar with those obtained in numeric relativity method \cite{NmRel0507,NmRel0511,NmRel0511103,tidalDeformationNumeric,NR2002,NR2004,NR2007,NR2018} and black hole perturbation theories \cite{BHPTwaveform1970,BHPTwaveform1971,BHPTwaveform1975,QNMcardoso0905,QNMcardoso1602,QNMcardoso1806}.

Section \ref{fullWFinnerstructure} will compare the relative motion orbits and gravitational wave forms following from our XOB method for three different time variation modes $z(t)$ of banana shape deformation. These three modes describe black holes with three typical shape deformation sensitivity or susceptibilities under the inhomogeneous back reaction force of gravitational wave radiation. By exhibiting the resultant gravitational wave forms' difference, we will show that the inner structure and the shape deformation sensitivity of black holes are observable experimentally instead of purely theoretical arguments. By assuming that gravitational wave carries no energy away, we will derive a universal upper bound for the late time quasi-normal mode's frequency as functions of the symmetric mass ratio of the binary system irrespective of their inner structure details. While by setting the conservative part of the XOB hamiltonian to the minimal possible value, we will get a lower bound for such quasi normal frequency. We will use existing data to test this two bounds quantitatively.

Section \ref{secConclusion} is the conclusion of the whole paper.

\section{The outside fixed position observer's description}
\label{secMetricOFPO}

The physical reality's probe dependence is a very widely accepted concept in high energy physics. For example, the hadron structures measured in the deep inelastic scattering experiments depends on the probe particle's momentum transfer. One purpose of this work is to show that, the horizon and singularity's observer independent is just a counter part of this concept in general relativity and black hole physics.

By the word of outside fixed position observer, we refer to all out side probe or detectors that are affected by the gravitation emanate from the to be black hole collapsing stars. These observers can only probe or detect physics related with the collapsing stars during their life time, i.e. existence duration of the probe themselves. Typical examples for these observers include, either member of a BHB system which are experiencing merger process, and/or all imaging light signals that are emitted from the background accretion disk and enter the detector of the event horizon telescope on earth. These observer or probe's positions are not truly fixed, but during their whole existence duration and in the spherical coordinate system of the being probed black holes, their radial positions are only slowly varying functions of the time coordinate which is understood as physic times by the infinitely far away investigators.

To these observers, the general form of spacetime metric of the collapsing star can be written as
\bea{}
&&\hspace{-5mm}ds^2_\mathrm{full}{=}-h{\cdot}B^{-1}dt^2{+}h^{-1}dr^2{+}r^2d\Omega^2, h{=}1{-}\frac{2GM}{r}
\label{metricOutObserverA}
\\
&&\hspace{-5mm}B^{-1}=1+h^{-2}\dot{M}^2/{M'}^2,M[t,r]^\mathrm{outside.matt.}_{occup.region}\equiv M_\mathrm{tot}
\label{metricOutObserverB}
\\
&&\hspace{-5mm}M[t,r]\xrightarrow[r<2GM_\mathrm{tot}]{t\rightarrow\infty}\frac{r}{2G}-\mathrm{small~deviation}
\label{asympMfunction}
\eea 
The function form of $B$ is determined by the normalisation of matter sources' four velocity $u^\mu=\frac{dx^\mu}{d\tau}=\{1,\frac{\dot{m}}{m'}\}\cdot\frac{dt}{d\tau}$. We choose the scaling of $t$ so that $hdt^2={d\tau}^2$, where $\tau$ is the proper time measured by observers fixed on the collapsing matter, up to a regular $r$-dependent scale factor. In the case of gravity dominating over all other interactions, neglecting the pressure is a rational doing. In this case, the Einstein equation $R_{\mu\nu}-\frac{1}{2}g_{\mu\nu}R=8\pi G\rho u_\mu u_\nu$ will tell us
\bea{}
\frac{\dot{m}\dot{m}'}{{m'}^2}{-}\big[\frac{2m}{r(r{-}2m)}{+}\frac{m''}{{m'}}\big]\frac{\dot{m^2}}{{m'}^2}
{-}\frac{(r{-}2m)m'}{r^2}{=}0
\label{einstEq0011}
\eea
\bea{}
\frac{\ddot{m}}{m'}{-}\big[\frac{3m}{r(r{-}2m)}{+}\frac{m''}{m'}\big]\frac{\dot{m}^2}{{m'}^2}
{-}\frac{(r{-}2m)(m{+}2rm')}{r^3}{=}0
\label{einstEq33}
\eea
where we used short notations $m=GM[t,r]$. Equation \eqref{einstEq0011} follows from the condition that $G^0_0G^1_1-G^0_1G^1_0=0$. Equation \eqref{einstEq33} follows from the condition of $G^\theta_\theta=G^\phi_\phi=0$ supplemented by \eqref{einstEq0011}. 

Eq\eqref{einstEq0011} can be written into the form of a first order differential equation and integrated formally
\bea{}
y'(t,r)-p(t,r)y(t,r)=q(t,r), y\equiv\dot{m}^2
\\
p\equiv\frac{2m''}{m'}{+}\frac{4m/r}{(r{-}2m)},q\equiv\frac{2(r{-}2m){m'^3}}{r^2}
\\
y=e^{\int^r_\mathrm{r_1}\!\!p(t,z)dz}{\cdot}c(t){+}\!\!\int^r_{\mathrm{r_2}}\!\!\!\!e^{\int^r_x\!p(t,z)dz}q(t,x)dx
\label{dotMintegExpression}
\eea
where $c(t)$ is an arbitrarily tunable function of $t$-coordinate, while $r_1$ and $r_2$ are two arbitrary reference point. Taking $\frac{m/r}{m'{-}1/2}$ as a slowly varying function of $r$, the integration $\int\!p(t,z)dz$ in \eqref{dotMintegExpression} can be done approximately so that
\bea{}
&&\hspace{5mm}\int\!p(t,x)
{\approx}[\log{m'^2}\!(x{-}2m)^{\frac{4m/x}{1{-}2m'_x}}]
\label{pApproxFormulation}
\\
&&\hspace{-8mm}\frac{\dot{m}^2}{m'^2}{\approx}{\!(r{-}2m)^{\!\frac{4m/r}{1{-}2m'_r}\!}}{\cdot}\big[c(t)
{+}\!\!\int^r_{r_2}\!\!\!
\frac{(x{-}2m)^{1{-}\frac{4m/x}{1{-}2m'_x}}2m'}{x^2}dx\big]
\label{approxdmsOpms}
\eea
This means that given initial mass function $m(0,r)$, we can always tune the form of $c(t)$ so that on the boundary of the matter occupation region $\frac{\dot{m}}{m'}|^\mathrm{m.o.r}_\mathrm{bndry}=\frac{dr}{dt}|^\mathrm{m.o.r}_\mathrm{bndry}$ matches with the geodesic motion of test particles freely released in the Schwarzschild metric caused by the total mass of the collapsing star. But according to equation \eqref{einstEq0011}, the initial value of $\dot{m}^2(0,r)$ is completely determined by the form of $m(0,r)$ thus allows no artificial tuning to implement a given $\frac{\dot{m}}{m'}(0,r^\mathrm{m.o.r}_\mathrm{bndry})$. This is because we neglect all pressure effects as well as the external force/potentials exerted on the matter contents so that their initial speed equals zero. In our numerical examples, we will simply set $c(t)=0$ and take $r_2$ as a small parameter to assure integral convergence and accuracy. In this case equation \eqref{approxdmsOpms} implies that as $m\rightarrow\frac{r}{2}$, $\frac{\dot{m}^2}{m'^2}\sim(r{-}2m)\rightarrow0$. In this limit, equation \eqref{einstEq0011} leads to
\bea{}
\dot{m}\sim(1{-}\frac{2m}{r})^{-1/2},{m'}\sim(1{-}\frac{2m}{r})^{-1},
\label{asympMfunctionSecondtime}
\eea
This explains the origin of asymptotic expression \eqref{asympMfunction}, which is in fact a result independent with respect to the matter source's equation of state. The Ricci scalar of spacetime metric \eqref{metricOutObserverA}-\eqref{metricOutObserverB} when simplified using \eqref{einstEq0011} and \eqref{einstEq33} has the following form
\beq{}
R=\frac{2(r-2m)^2{m'}^3}{r^2[(r-2m)^2{m'}^2+r^2\dot{m}^2]}
\eeq
This is regular as $t\rightarrow\infty$, because the asymptotic behaviour of \eqref{asympMfunctionSecondtime}. Divergent happens only when we try to go beyond the Future of infinite Far Future (FiFF), that is, when we use alternative time coordinate which covers range outside that $t$ does.

Eqs\eqref{einstEq0011}-\eqref{einstEq33} allows us to express all the higher order time derivatives of $m$ in terms of its spatial derivatives, e.g.
\beq{}
\ddot{m}{=}\big[\frac{3m}{r(r{-}2m)}{+}\frac{m''}{m'}\big]\frac{\dot{m}^2}{m'}
{+}\frac{(r{-}2m)(m{+}2rm')m'}{r^3}
\eeq
\bea{}
\dddot{m}{=}\dot{m}^3\big[\frac{m'''}{{m'}^3}{+}\frac{9mm''}{r(r{-}2m){m'}^3}{+}\frac{4r^2m'{-}6rm{+}24m^2}{r^2(r-2m)^2{m'}^2}\big]
\\
{+}\dot{m}\big[\frac{2[6m^2{-}rm(1{-}8m'){+}3r^2{m'}^2]}{r^4}{+}
\nonumber
\\
\frac{3(r{-}2m)(m{+}3rm')m''}{r^3m'}\big]
+\frac{(r{-}2m)^2{m'}^3(2m{+}3rm')}{r^5\dot{m}}
\nonumber
\eea
These expressions tell us that, as $t\rightarrow\infty$ the $n$-th order time derivative of $m$ are proportional to $(r-2m)^{-n}\sim\infty$. This makes numeric integration of equations \eqref{einstEq0011} or \eqref{dotMintegExpression} more and more difficult as we approach the configuration $m(t,r)\approx\frac{r}{2}$. However, the basic feature of $\frac{dr}{dt}|_\mathrm{fix.mss.pt}=\frac{\dot{m}}{m'}\sim(r-2m)^\frac{1}{2}\rightarrow0$ is very robust. This causes the outside probes not being able to probe $m(t,r)$' evolution into profiles of $r-2m<0$ in any finite $t$ time. Physically this is just the fact that, as outside observers or probes, they cannot see the formation of horizons in any finite $t$ durations. Of course, if the proper time description is adopted, the co-moving observers can see matter's collapse into the horizon in finite durations, just as was illustrated in references \cite{dfzeng2017}. However, since those observers or detectors are freely falling, they will experience not any exotics as they pass through the horizon surface.
\begin{figure}[h]
\includegraphics[totalheight=50mm]{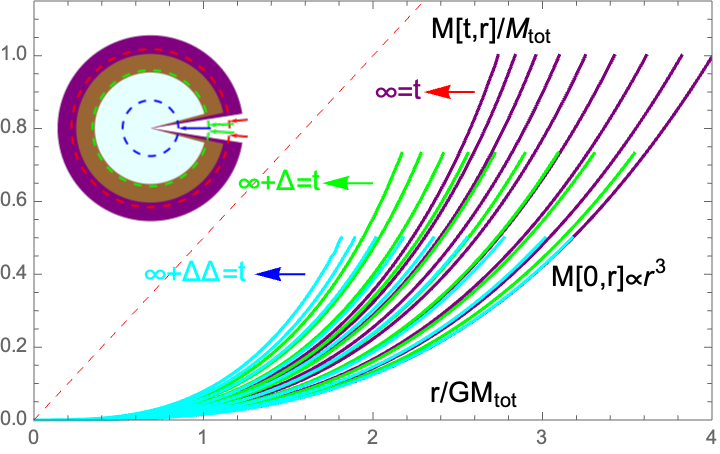}
\includegraphics[totalheight=50mm]{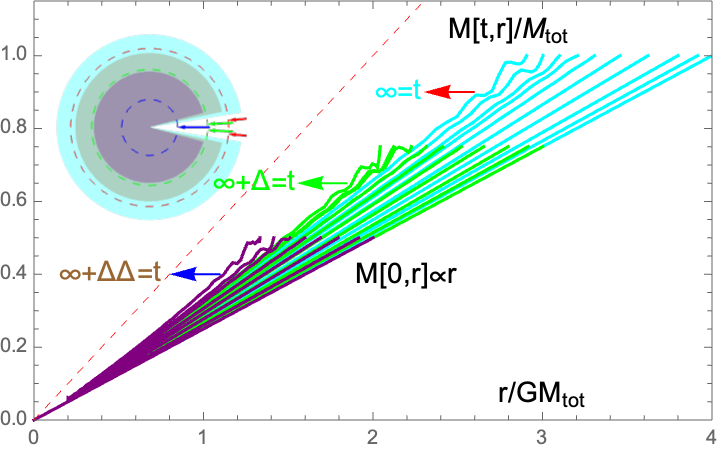}
\caption{The $t$-time evolution of the mass function of two collapsing stars. The upper has central sparser initial configuration, the downer has a central denser one. Different colour curves family denotes the mass function evolution inside three concentric spheres of different radius.}
\label{figEvMassProfOutsideObserver}
\end{figure}
\begin{figure}[h]
\includegraphics[totalheight=50mm]{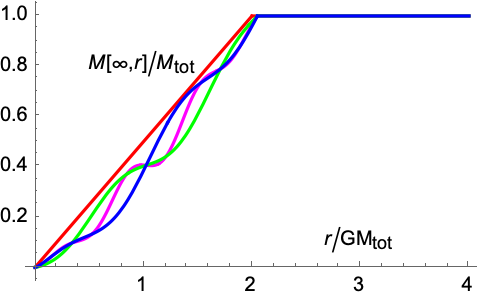}
\caption{The red line is the asymptotic mass function of all collapsing stars after infinite $t$-time evolution. $t$ is the time of the outside fixed position observers. The other three curves are the mass function of collapsing star after finite $t$-time evolution. The matters live in the region $0{<}r{<}2GM_\mathrm{tot}{+}\epsilon$.}
\label{figMassProfOutsideObserver}
\end{figure}
\begin{figure}[h]
\includegraphics[totalheight=45mm]{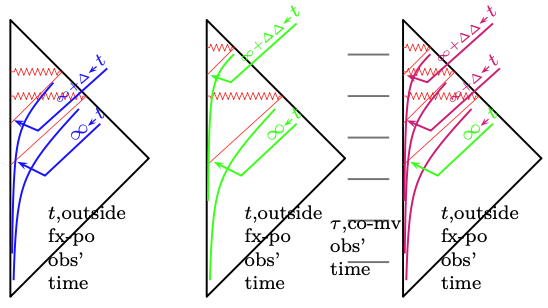}
\caption{The Penrose-Carter diagram of the three inner mass profiles of FIG.\ref{figMassProfOutsideObserver}. The horizon and singularity will be realised only when $\infty<t$. The two blue hyperbolas in the leftmost sub-figure correspond to the $r=2GM_\mathrm{tot}$ and the equal-$r$ sphere which are mostly close to the $M=r/2G$ line. The two green hyperbolas in the middle and the three maganta hyperbolas in the rightmost are similar.}
\label{figPCfxoInnStructABC}
\end{figure}

Consider two simple initial mass function (i) $m(0,r<1)=\frac{1}{4}r^3$, (ii) $m(0,r<1)=\frac{1}{4}r^1$ and $m(0,1<r)=1$. The former has central sparser initial mass density profile $\rho\sim(1-\frac{r^2}{4})^{-1}$, the latter has a central denser and singular profile $\rho\sim r^{-1}$ but no horizon apriori. For these two initial mass functions the $t=0$ integration \eqref{dotMintegExpression} can be done exactly as well as analytically
\bea{}
&&\hspace{-6mm}\mathrm{(i)}~\dot{m}^2(0,r)=\frac{r^4c(t)}{(2{-}r^2)}{+}\frac{9r^4}{128}\big[\frac{(2-r_2^2)^3}{2-r^2}{-}(2-r^2)^2\big]
\\
&&\hspace{-6mm}\mathrm{(ii)}~\dot{m}^2(0,r)=r^2c(t)+\frac{r^2}{128}\big[\frac{1}{r_2^2}-\frac{1}{r^2}\big]
\eea
FIG.\ref{figEvMassProfOutsideObserver} displays their latter time evolution following from equations \eqref{einstEq0011}-\eqref{einstEq33}. From the figure we can easily see that different radius concentric shells need different $t$-duration to reach their horizon size. The duration difference $\Delta$ is finite. This means that to the outside detectors, the horizon of a collapsing star is not a single equal-r surface of $r=2GM_\mathrm{tot}$, but the collection of all concentric spheres of different sizes $\{r_i=2GM^i\}$, where $M^i$ means the mass of matters inside the sphere of radius $r_i$ and $i$ varies continuously on classical levels. Each one of those horizons is never reached but only infinitely closely reachable by the corresponding shells at different $t$-time epochs depending on the initial mass function of the star. As outside observer or probes, one can use the initial mass function to represent the microscopic state of the collapsing star with approximate horizon of radius $r_h=2GM_\mathrm{tot}+\epsilon$. FIG.\ref{figMassProfOutsideObserver} gives an illustration for three typical examples, while FIG.\ref{figPCfxoInnStructABC} is their Penrose-Carter diagram representation.

The physical black holes form through gravitational collapses are defined by their asymptotic horizon. The asymptotic horizon is only mathematical surface which is infinitely approachable but never reached in any finite physic time. It provides us a boundary to look the physical black holes as statistic systems and introduce parameters such as internal energy and entropy to describe their macroscopic features. Just as all normal statistic systems do, the microscopic state of the physical black holes can be simply attributed to the mechanical motion of their matter contents. At here, this motion is the freely falling of concentric matter shells consisting of the collapsing star under randomly specified initial radius and radial speed. The randomness of initial conditions arise from the uncertainty principle. We will discuss the quantisation of this mechanical motion and show that its wave functionals are indeed degenerated in the number of ways required by the area law formula of Bekenstein Hawking entropy. But before this doing, let us consider this motion's description from the inside co-moving observer's aspect.

\section{The inside co-moving observer's description}
\label{secMetricComObserver}

By the term inside co-moving observer, we refer to all local detectors or probes which are freely falling together with the matters consisting of the collapsing star and use proper time as time coordinate to describe physic evolutions accessible to them. As local probes, these observers are not necessarily point particles. As long as their size is much smaller than the space-time region's scale they are detecting, things will be okay. Their detection are just interactions between them and the to be detected spacetime region itself. The interaction between different parts of the system is also a detection which has the potential of putting the microscopic state of the whole system onto an ergodic evolution orbit. To these observers, the most general metric ansatz for an isotropic but inhomogeneous collapsing star can be written as
\beq{}
ds^2_\mathrm{in}{=}{-}d\tau^2{+}\frac{\big[1{-}\big(\frac{2GM}{\varrho^3}\big)\!^\frac{1}{2}\!\frac{M'\!\varrho}{2M}\tau\big]^2\!d\varrho^2}{a[\tau,\varrho]}{+}a[\tau,\varrho]^2\varrho^2d\Omega^2_2
\label{genOSmetric}
\eeq
\beq{}
ds^2_\mathrm{out}=-d\tau^2{+}\frac{r_s^{2/3}d\varrho^2}{[\frac{3}{2}(\varrho{-}\tau)^\frac{2}{3}]}{+}[\frac{3}{2}(\varrho{-}\tau)^\frac{2}{3}]^2r_s^\frac{2}{3}d\Omega_2^2
\label{outMetric}
\eeq
where as was done in the previous section, we have neglected the pressures so that the inner structures of the system are characterised by a single function $M[\varrho]$ exclusively. This is reasonable when the gravitation dominates all non-gravitational exclusive  interactions; $\tau$ and $\varrho$ are so called Lemaitre coordinate and
$r_s\equiv2GM_\mathrm{tot}$. By the conventional Boyer-Linderquist coordinate, $ds^2_\mathrm{out}$ can be equivalently written as 
\beq{}
ds^2_\mathrm{out}{=}{-}hdt^2{+}h^{-1}dr^2{+}r^2d\Omega^2,h{=}1{-}\frac{2GM_\mathrm{tot}}{r}
\eeq
The advantage of Lemaitre coordinate is that, we can more directly see the smoothness of the inside-outside metric's connection. The advantage of Boyer-Linderquist coordinate is that, we can more easily understand what we are describing is a simple neutral and spherical symmetry black hole in the standard general relativity.  Since we neglect all exclusive pressures, the concrete form of $a[\tau,\varrho]$ can be obtained exactly from the dust sourced Einstein equation  $R_{\mu\nu}-\frac{1}{2}g_{\mu\nu}=8\pi G\{\rho,0,0,0\}$
\bea{}
a[\tau\!\in\!|_0^{\!\frac{p\;\!\!^\varrho}{4}},\varrho]\!=\!\big(1\!-\!\frac{4\tau}{p^\varrho}\big)^{\!\frac{2}{3}}
,~a[\tau|_{\frac{p\;\!\!^\varrho}{4}}^{\frac{p\;\!\!^\varrho}{2}},\varrho]{=}{-}a[\frac{p\;\!\!^\varrho}{2}{-}\tau,\varrho]
\rule{2mm}{0pt}
\label{genOSscalefactor}
\\
a[\tau|_{\frac{p\;\!\!^\varrho}{2}}^\frac{p\;\!\!^\varrho}{~},\varrho]{=}{-}a[p^\varrho{-}\tau,\varrho]
,~
a[\tau|_{p\;\!\!^\varrho}^{p\;\!\!^\varrho{\!\scriptscriptstyle+}},\varrho]=a[\tau{-}p^{\varrho},\varrho]
\rule{2mm}{0pt}
\\
p^\varrho\equiv\frac{8}{3}\big(\frac{\varrho^3}{2GM[\varrho]}\big)\!^\frac{1}{2}
,~M[\varrho\!\geqslant\!\varrho_\mathrm{max}]{=}M_\mathrm{tot}
\rule{13mm}{0mm}
\label{aperiodic}
\eea
Since $a(\tau,\varrho)$ is negative allowable, the sign of the coefficient of the $d\varrho^2$ term in \eqref{genOSmetric} must be defined in such a way that the whole term is positive definitely. 

\begin{figure}[ht]
\includegraphics[totalheight=40mm]{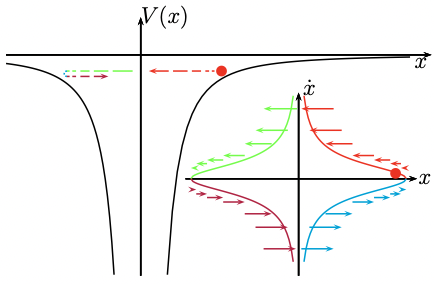}
\includegraphics[totalheight=25mm]{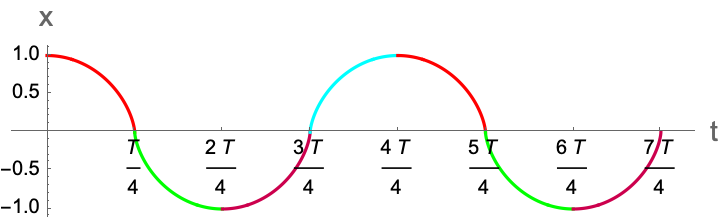}
\caption{The upper part is the potential and phase diagram of oscillators driven by a linearly inverse potential or square inverse force, the lower part is its oscillation function.}
\label{figSqInversePotential}
\end{figure}

For each given $\varrho$, the behavior of $a[\tau,\varrho]$ involved in \eqref{genOSmetric} or \eqref{genOSscalefactor} is just the oscillation driven by a square inverse force or linearly inverse potential, 
\beq{}
m\ddot{x}+\frac{k}{x^2}=0~\mathrm{or}~\frac{1}{2}m\dot{x}^2{-}\frac{k}{x}{=}\varepsilon
\label{exoticOscillator}
\eeq
The full period of this oscillation is $T{=}\frac{k \sqrt{m}}{\sqrt{2} (-\varepsilon)^{3/2}}\frac{\pi}{2}$. During each first quater period, the oscillation equation can be integrated implicitly
\bea{}
&&\hspace{-9mm}
t(x)=\frac{k \sqrt{m}\arcsin[(1+\frac{x\varepsilon}{k})^\frac{1}{2}]}{\sqrt{2} (-\varepsilon)^{3/2}}
-\frac{\sqrt{m x (k+x \varepsilon)}}{\sqrt{2}\varepsilon}
\label{xoftInverseFucntion}
\eea
and the full oscillation function can be written as
\bea{}
&&\hspace{-5mm}\{t,x\}\equiv\left\{
\begin{array}{ll}
\{~0+t(x),+x\},&0<t\leqslant\frac{T}{4}
\\
\{\frac{T}{2}-t(x),-x\},&\frac{T}{4}<t\leqslant\frac{T}{2}
\\
\{\frac{T}{2}+t(x),-x\},&\frac{T}{2}<t\leqslant\frac{3T}{4}
\\
\{T-t(x),+x\},&\frac{3T}{4}<t\leqslant T
\end{array}
\right.
\label{xoftContinuated}
\eea
FIG.\ref{figSqInversePotential} displays the potential, phase diagram and $x\mathrm{vs}.t$ curve of this oscillation in details. By the standard Boyer-Lindquist to Lemaitre coordinate transformation
\bea{}
&&\hspace{-5mm}dt+\sqrt{\frac{r_s}{x}}\big(1{-}\frac{r_s}{x}\big)^{-1}dx\equiv d\tau
\\
&&\hspace{-5mm}dt+\sqrt{\frac{x}{r_s}}\big(1{-}\frac{r_s}{x}\big)^{-1}dx\equiv d\varrho
\\
&&\hspace{-5mm}x=\big[\frac{3}{2}(\varrho{-}\tau)\big]^\frac{2}{3}r_s^\frac{1}{3}
\eea
the oscillation function \eqref{xoftInverseFucntion}-\eqref{xoftContinuated} will become the function \eqref{genOSscalefactor}-\eqref{aperiodic} with fixed $\varrho$ routinely.

\begin{figure}[ht]
\includegraphics[totalheight=65mm]{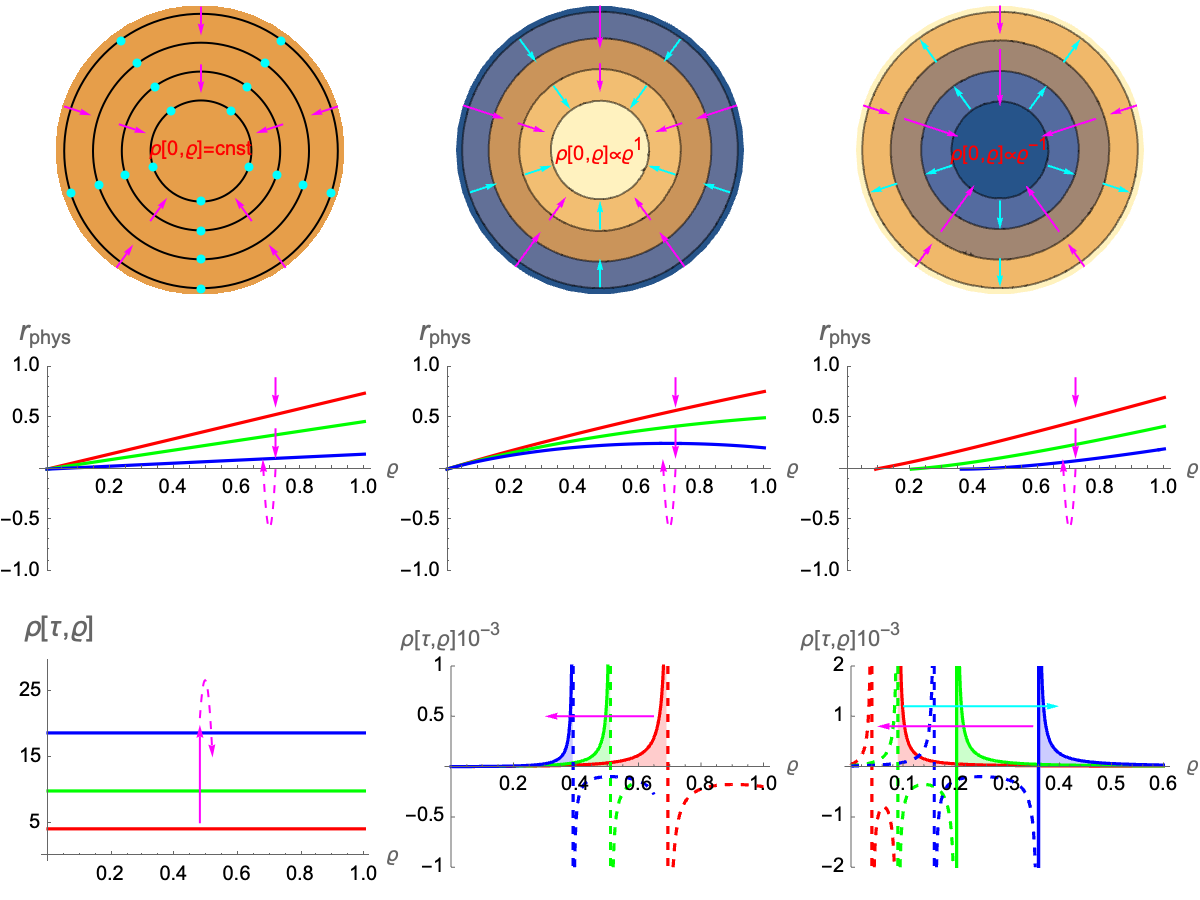}
\caption{The evolution of three representative dust stars as the $\tau$-time varies. The magenta arrows denote the co-moving grid's contraction as a whole. The left has uniform initial mass density. The cyan dot displays the matters' fixing on the co-moving grid. The middle has a central-sparser initial mass density $\rho[0,\varrho]\sim\frac{dM[\varrho]}{4\pi\varrho^2d\varrho}=\varrho^1$. The matters' migration on the co-moving grid, marked by cyan arrows, is from the superficial to central region so that the co-moving $\varrho$-coordinate covers smaller and smaller regions. The right has a central-denser initial mass density $\rho[0,\varrho]\sim\varrho^{-1}$. The matters migration on the co-moving grid in this case is from the central to the outer region, marked by cyan-arrows. } 
\label{figOCOclassic}
\end{figure} 

The mass function $M[\varrho]$ involved in \eqref{genOSmetric} and \eqref{genOSscalefactor}-\eqref{aperiodic} are completely arbitrary at the classic level. Its physics meaning is the initial mass function of the collapsing star on the co-moving coordinate $\varrho$ line. FIG.\ref{figOCOclassic} displays the $\tau$-time evolution of three representative collapsing stars characterised by different $M[\varrho]$. In the left most uniform initial density case, the evolution of the system is completely determined by the contraction and expansion of the co-moving grid as a whole. But in the other two cases, the co-moving grid's contraction and the matter's migration on the grid are asynchronous. In the central sparser case of the middle column, the matters' migration on the co-moving grid is from the superficial layers to the central ones, so that the matters are distributed on smaller and smaller co-moving grids. While in the central-denser case of right most column, the matters' migration on the co-moving grid is from the central region to the superficial layers, so that they cover larger and larger co-moving grids. In both latter cases, the matters' migration on the co-moving grid is towards a more and more homogeneous configure until they reach an exactly uniform distribution and over-cross each other. At the over-cross epoch, the physical mass density $dM/(a^3\varrho^2d\varrho)$ is divergent, but the density $dM/\varrho^2d\varrho$ on the co-moving grid is finite. After that epoch, the mass-energy distribution will retrieve its initial configuration but with the anti-podal point interchanged. So the matters' gathering onto the central point and causing divergence there is not the terminal of their motion inside the horizon but a normal epoch of a periodical over-cross oscillation. Of course this oscillation happens outside the domain of time definition of the outside observers. To them this oscillation especially the matter shells' expansion away from the gravitational central happens only in a quantum parallel universe or statistic ensemble spacetime which is causally in-connected with the one they own live in and is understandable only as the results of the quantum fluctuation implied by the uncertainty principle.

Relative to other mono-directional collapsing star solutions to the sourceful Einstein equation such as  Oppenhiemer-Snyder \cite{oppenheimer1939,MTWbook}, Yodzis-Seifert-Muller \cite{Yodzis1973,Yodzis1974} and Vaidya \cite{Vaidya1950,Vaidya1970} in the history, the most important feature of our oscillation solutions family above is its not taking the dusts' collapse to $a[\tau,\varrho]=0$ and hitting on the singularity as the terminal of their motion inside the horizon. Instead it takes such event only as an intermediate epoch of the dusts' motion inside the horizon. The next step and following on motion is their passing through each other and periodic oscillation across the central point. Their motion over cross the central point is an analogue of two classic wave's superposition and independent propagation after encountering each other. This over cross oscillation mechanism is firstly noted and quantised with a functional Schrodinger equation to resolve the Schwarzschild singularity in reference \cite{dfzeng2017} and then quantised with the method of shell-decomposition in references \cite{dfzeng2018a,dfzeng2018b} to interpret the origin of Bekenstein-Hawking entropy and resolve the black hole information missing puzzle caused by hawking radiation. While in reference \cite{dfzeng2020}, exact solutions for the simple $AdS_{2+1}$-Schwarzshild are provided and the corresponding Bekenstein-Hawking entropy formula is analytically proved by the integer number partition formula.

\begin{figure}[ht]
\includegraphics[totalheight=48mm]{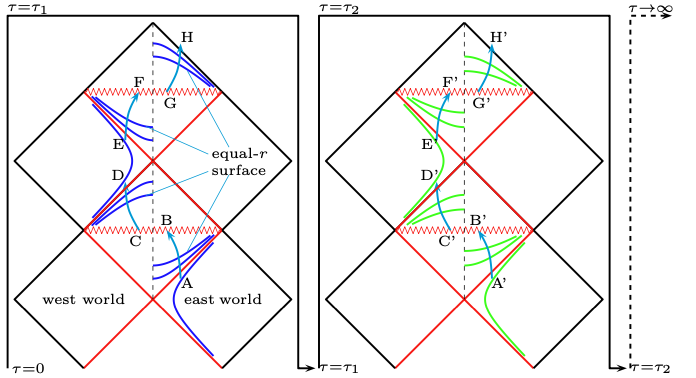}
\caption{To a co-moving local observer which uses proper time $\tau$ to describe physic evolutions, the Schwarzschild singularity caused by gravitational collapse is trivially traversable as the horizon is. So its world line in the ${\tau,\rho}$ plane A,B,C,D$\cdots$,A\!$'$,B\!$'$,C\!$'$,D\!$'$ is periodic. The Schwarzschild singularity represents not any terminal of physic evolution. But the definition domain of proper time $\tau$ covers that of the outside fixed position observers' time $t$ infinite times. During the infinitely long $\tau$-time evolution, the shell structure a freely falling co-moving observer see would not be exactly periodic due to the many-body essence of the interactions among the different shells. This would cause ergodicity of the shell structure's evolution.} 
\label{figPCcstarMultCover}
\end{figure} 

The existence of our solutions family \eqref{genOSmetric}-\eqref{aperiodic} does not contradict with the singularity theorem of Penrose and Hawking \cite{Penrose1965,Hawking1976,geroch1979}, because all matters described in this solutions family indeed fall onto the central point and cause a spatial-like singularity in finite proper time as expected. To the co-moving observer, such a singularity is nothing but an equal-$\tau$ hyper-surface. Just like the horizon can be trivially traversed, the spatial-singularity which occupies only a future point on the co-moving observer's world line is also trivially traversable in almost no $\tau$-time, after which all kinematic parameters of the system get their $\tau$-time reversed value routinely. Tidal forces affects only the finite sized objects. However, just as we pointed out in the introduction section, such effects are time inverse symmetric. So even when the finite sized objects are involved in a physic collapsing process, the status information encoded in such objects will be retrieved as the singularity is traversed. This means that we can define the microscopic state of the black hole with the initial status of its consisting matters. For an exactly spherical symmetric Schwarzschild black holes, the only possible initial status information is its matter contents' layering structure and each layer's radial collapsing speed. 

To the outside fixed position observer or physical probes, the asymptotic horizon provides an ideal excuse for them to say that the system has to be considered as an ensemble of many parallel universe, each contains a collapsing star with equal mass but different initial radial mass profile. To the freely falling co-moving observers, this excuse cannot be used. However, the fact that the domain of proper time $\tau$'s definition covers that of the coordinate time $t$ many times brings us another interpretation, ergodicity. That is, during the infinitely long $\tau$-time evolution, the point representing the microscopic state of the black hole will walk through each point of the phase space that represents all possible microscopic states of the collapsing star. See FIG.\ref{figPCcstarMultCover} for references. To display the oscillations inside the horizon more conveniently, we extended the rule of conventional Penrose Carter diagram so that the east world and west world of the spacetime are simultaneously displayed in the $\{\tau,\rho\}$ plane. We introduced this diagram method firstly in reference \cite{dfzeng2021}. We will show in the following that the microscopic state definition based on the ergodicity of co-moving observers and that based on the uncertainty principle of the outside fixed position observers are equivalent to each other.

In both FIG.\ref{figPCfxoInnStructABC} and FIG.\ref{figPCcstarMultCover}, the horizon represented by red straight lines and the singularity represented by red zigzags are physical realities only to the investigators which can see the Future of infinitely Far away Future (FiFF), we will call the viewpoint of such observers as global of God-type. While the viewpoint adopted by the true physic probes will be called local.  Both the outside fixed position observer and the freely falling co-moving observers are local. To them the horizon and singularity are only mathematical realities instead of physical ones, because they have not any physical ways to detect such realities' existence.

\section{Quantisation}
\label{secQuantisation}
  
As the definition function for black hole microscopic state, both the asymptotic $M[\infty,r]$ in \eqref{asympMfunction} and the co-moving mass profile $M[\varrho]$ in \eqref{genOSmetric}-\eqref{aperiodic} at the classic level are arbitrary thus un-countable. However, at the quantum level, things are rather different. We have two methods to quantise the system canonically. The first is to use the hamiltonian formulation of general relativity to build a Wheeler-Dewitt like functional Schrodinger equation which takes the instantaneous mass function $m(t,r)$ as independent variable and the radial mass-energy density profile $\rho[t,r]$ as eigenvalues. 
\beq{}
\big[\frac{i^2\hbar^2\delta^2}{\delta m(r)^2}+\frac{r^232\pi G_{\scriptscriptstyle\!N}\rho}{m'h^2}+0\big]\Psi[m(r)]=0
\label{funcSchrodingerEq}
\eeq
where $h\equiv1-\frac{2m(r)}{r}$ and $m(r)$ is the value of $m(t,r)$ at any given time $t$. The $3+1$ decomposition or foliation of the spacetime itself defines the relation between $m(t,r)$ and the instantaneous energy density $\rho(t,r)$. 
\beq{}
h^{-\frac{1}{2}}r^2\big[-\frac{2m'}{r^2}+8\pi G_{\scriptscriptstyle\!N}\rho\big(h^{-2}\frac{\dot{m}^2}{m'^2}+0\big)\big]=0
\label{classicHamiltonianConstraint}
\eeq
In the classic hamiltonian constraint $m$ should be understood as the function of $t$ and $r$ simultaneously, while in its quantum version $m$ is understood as the function of $r$ only. This quantisation method is a generalisation of the minimal super space quantum cosmology. We firstly proposed this generalisation in reference \cite{dfzeng2017} but with an error happens in the functional Schrodinger equation corrected in \cite{dfzeng2021}. In their counter parts of references \cite{dfzeng2017,dfzeng2021}, the zero term positions of \eqref{funcSchrodingerEq} and \eqref{classicHamiltonianConstraint} are occupied by terms of $\rho^2$ and $1$ respectively. This difference is due to the time coordinate adopted in this work to define the hamiltonian constraint is the outside fixed position observers time $t$ instead of the proper time $\tau$ of the inside co-moving observers.

To directly discretise the independent variable $m(r)$ and resolve the functional Schrodinger equation \eqref{funcSchrodingerEq} is very difficult, see reference \cite{dfzeng2017} for our earlier attempts. Beginning from reference \cite{dfzeng2018a} and through out references \cite{dfzeng2018b,dfzeng2020,dfzeng2021,dfzeng2022}, we proposed a new discretisation stratagem. In this new stratagem, we decompose the whole collapsing star into many concentric shells $\{m_i\}$ and quantise each of them canonically. The shell decomposition here is non-apriori. All schemes allowed by the following two conditions would be considered,
\bea{}
&&\hspace{-5mm}m_1+m_2+\cdots+m_k=M_\mathrm{tot}
\label{massSumRule}
\\
&&\hspace{-5mm}\frac{GM_im_i}{(2-2\gamma_i^2)^\frac{1}{2}\hbar}=n_i=1,2,3,\cdots
\label{radialQuantizeCondition}
\eea
The former follows from the total mass sum rule, while the latter follows from the square integrability of $m_i$'s wave function; $M_i\equiv\sum\limits_{i'\leqslant i} m_{i'}$; $\gamma_i$ or $n_i$ is the radial quantum number of shell $m_i$'s wave function and classically they are determined by the shell's initial radius. Quantum mechanically the mass function of the whole collapsing star can be written as
\beq{}
M[r]=\sum_im_i|\psi_i(r)|^2
\label{mfuncQuDefinition}
\eeq 
since $\psi_i(r)$ here is the eigenfunction of the hamiltonian whose conjugating time is that of the outside fixed position observers, this quantum mass function corresponds to the asymptotic function $M[\infty,r]$ of eq\eqref{asympMfunction}. This can also be taken as the initial mass function on the co-moving grid since the initial time epoch on the co-moving grid is arbitrarily choosable. 

Obviously, the mass function's quantum definition \eqref{mfuncQuDefinition} tells us that it is the combination of the shell decomposition and initial radius assigning scheme that defines the microscopic state of the collapsing star, neither one of them alone is enough to do that work. So our number counting of the microscopic state of the system should be made on
\beq{}
\#\mathrm{{\bf\!p}artition}\{m_i\}\otimes\mathrm{{\bf\!e}xitation}\{n_i\}
\eeq
the partition here is order-concerned, e.g. $\{m_i\}=\{1,0.7,0.3\}$ and $\{m_i'\}=\{1,0.3,0.7\}$ are two different partitions of $M_\mathrm{tot}=2$. Two attentions must be payed with the number counting. The first is, the global maximal value position of each shell's wave function modular  $x^{\psi_i}_\mathrm{max}$ must happen outside $r^i_h=2G\sum\limits_{i'=0}^im_{i'}$ but not too far away from it. $r^i_h$ is the classic horizon defined by the masses the shell $m_i$ wraps. ``not too far'' means that the global maximal position of the outmost shell's wave function should take the lowest value which lies outside the classic horizon $r_h=2GM_\mathrm{tot}$. The second is, any radial excitation scheme should not make the global maximal value of shell $m_i$'s wave function $|\psi_i^2|$ happens outside that of the shell $m_{i+1}$, for all $i$. The position of these maximal values corresponds to the classic asymptotic radius of the shells. This implies that the the radial excitation number $\{n_i\}$ cannot be set arbitrarily. It can be numerically shown that given shell mass decomposition $\{m_i\}$, the number of possible $\{n_i\}$-values' assigning schemes which satisfy these two requirements grows at most polynomially with the total mass $M_\mathrm{tot}$ increases. This polynomial non-uniqueness is the origin of logarithmic corrections to the area law formula in our microscopic state definition for black hole.

From the viewpoint of the outside fixed position observers, the classic motion of each shell $m_i$ is freely falling in an effective Schwarzschild geometry determined by the total mass of shells inside $m_i$ (including $m_i$ itself)
\bea{}
&&\hspace{-5mm}ds^2=-h_idt^2+h_i^{-1}dr^2+r^2d\Omega^2
\label{effMetricmi}
\\
&&\hspace{-5mm}h_i=1-\frac{2GM_i}{r}
,~
M_i=\sum_{i'=1}^{i}m_{i'}
\eea
Using $x$ and $x^\mu(\lambda)\equiv\{t(\lambda),x(\lambda),0,0\}$ to denote the radius of shell $m_i$ and the world line of a representative point's moving on it, the standard geodesic equation and four velocity normalisation of the representative point tell us
\bea{}
&&\hspace{-5mm}\ddot{t}+\Gamma^{(i)t}_{\mu\nu}\dot{x}^\mu\dot{x}^\nu=0\Rightarrow h_i\dot{t}=\gamma_i=\mathrm{const}
\label{geodesicEomZeroComponent}
\\
&&\hspace{-5mm}-h_i\dot{t}^2+h^{-1}_i\dot{x}^2=-1
\label{geodesicEomFourVelocNormalisation}
\eea
where $\Gamma^{(i)\sigma}_{\mu\nu}$ and $\gamma_i$ are the usual Christoffel symbol corresponding to the effective metric \eqref{effMetricmi} and an integration constant of the geodesic equations respectively. $\gamma_i=0$ corresponds that the shell is released from $r_{ini}=2GM_i$ and $\dot{r}_{ini}=0$; $\gamma_i=1$ corresponds that the shell is released from $r_{ini}=\infty$ and $\dot{r}_{ini}=0$. The former implies that the surface $r=r_\mathrm{ini}$ happens to be a horizon. The latter implies the masses inside $r=r_\mathrm{ini}$ is far from forming horizon. From the two equations of \eqref{geodesicEomZeroComponent} and \eqref{geodesicEomFourVelocNormalisation}, we can easily derive out
\beq{}
\gamma_i^2-\dot{x}^2-h_i=0
\label{constraintA}
\eeq
Multiplying both sides of this equation with the shell mass $m_i$ and manipulate the resultant equation routinely, we will get
\beq{}
m_i\dot{x}^2-\frac{GM_im_i}{x}-m_i(\gamma_i^2-1)=0
\label{constraintB}
\eeq
This is nothing but the hamiltonian constraint of the shell's motion. 

Now replace the classic momentum $m\dot{x}$ into its operator form $i\hbar\partial_x$ and introduce a wave function $\psi_i(x)$ to denote the probability amplitude the shell $m_i$ be measured at radius being $x$ status, we will get the quantum version of the Hamiltonian constraint \eqref{constraintB}
\beq{}
[-\frac{\hbar^2}{2m_i}\partial_x^2-\frac{GM_im_i}{x}-m_i(\gamma_i^2-1)]\psi_i(x)=0
\label{eigenStateSchrodinger}
\eeq
This quantises the motion of shell $m_i$ canonically. All other shells can be quantised similarly. Directly multiplying their wave-functions together, we will get the wave functional of the whole collapsing star as follows
\begin{align}
\Psi[M(r)]=\psi_0\otimes\psi_1\otimes\psi_2\cdots, \sum_im_i=M_\mathrm{tot}
\label{directProductWaveFunction}
\end{align}
Equation \eqref{eigenStateSchrodinger} is almost the standard eigenstate Schrodinger equation with a coulomb like potential except the normalisation condition
\beq{}
\int_0^\infty|\psi_i|^2dx=1
\label{wfNormalisation}
\eeq
which involves no angular space factor $4\pi x^2$ because $\psi_i(x)$ is the probability amplitude of the spherical shell as a whole to lie on the state of radius taking $x$ value instead of the probability amplitude a point particle to be measured at position $(x,\theta,\phi)$. 
The eigenfunction and eigenvalue of \eqref{eigenStateSchrodinger}  can be written down as
\bea{}
&&\hspace{-7mm}\psi_i=N_ie^{-\bar{x}}\bar{x}L_{n_i-1}^1(2\bar{x}),\bar{x}\equiv m_ix(2\mm2\gamma_i^2)^\frac{1}{2}/\hbar
\label{singleShellWvfunction}
\\
&&\hspace{-7mm}m_i(\gamma_i^2-1){\equiv}E_i=-\frac{(GM_im_i)^2m_i}{2n_i^2\hbar^2}
\label{enQuantizConditionSchwz}
\eea
where $n_i=1,2,3\cdots$ take positive integer values and $L_{n_i-1}^1(2x)$ is the associated Lagurre polynomial; $N_i$ is the normalisation factor following from \eqref{wfNormalisation}. Equation \eqref{enQuantizConditionSchwz} explains the origin of our microscopic state number counting constraint \eqref{radialQuantizeCondition}.

Although the radial quantisation condition \eqref{enQuantizConditionSchwz} or \eqref{radialQuantizeCondition} for each component shell provides countability for the microscopic state of the system, the number of state counted basing on equations \eqref{massSumRule}-\eqref{radialQuantizeCondition} is still infinite. This is because both the two conditions do not impose discreteness on the masses $\{m_i\}$ of the component shells, but only on the quotients $\{\frac{GM_im_i}{(2-2\gamma_i^2)^{1/2}\hbar}\}$ and $\gamma_i$ are not required to be zero exactly. As a regularisation for this countable infinite, we introduce the following concept of partitions' distinguishability
 \bea{}
  &&\hspace{-8mm}two~partitions~of~equal~number~of~shells~\{m_i\}\&
 \nonumber  
 \\
 &&\hspace{-8mm}\{m'_i\}~are~indistinguishable~on~precision~\varepsilon~if
 \label{distinguishabilityRule}
 \\
 &&\hspace{-8mm}{\scriptstyle\Delta}m_i\equiv|m_i-m'_i|\leqslant\min\{\frac{\varepsilon}{GM_{i}},\frac{\varepsilon}{GM'_{i}}\}~for~all~i;
 \nonumber
 \\
 &&\hspace{-8mm} {two~partitions~of~unequal~shell~numbers}
 \nonumber\\
 &&\hspace{-8mm} {are~always~distinguishable}
 \nonumber
 \eea
 By this standard, we can quantitatively state that to precision $\varepsilon=1$, the finest way of decomposing the collapsing star into concentric shells is that defined by the condition that for all $i$, $\frac{GM_im_i}{(2-2\gamma_i^2)^{1/2}\hbar}=1$ and $\gamma_i=0$.  We will call this decomposition scheme as fundamental. By this fundamental scheme, the number of shells the collapsing star can be partitioned into is $k\approx\frac{GM_\mathrm{tot}^2}{2\sqrt{2}}$. This is because
\bea{}
&&\hspace{8mm}GM_i(M_i-M_{i-1})=\sqrt{2}\Rightarrow
\\
&&\hspace{-7mm}M_i\xrightarrow[enough]{i~large}\big(\frac{2\sqrt{2}i}{G}\big)^\frac{1}{2},i^\mathrm{max}_{\mathrm{fix}M_\mathrm{tot}}\approx\frac{GM^2_\mathrm{tot}}{2\sqrt{2}}\equiv k
\eea

For the fundamental shell decomposition above, to assure that the global maximal value of each shell $m_i$'s wave function modular square happens outside the classic horizon $r^i_h\equiv2GM_i$, the lowest level radial quantum number assigning scheme is to set $n_i=2$ for all $i$. This is because in this case, the global maximal value of $|\psi^2_{n_i}(x)|$ happens on $x^{\psi_{n_i}}_\mathrm{max}=\frac{3+\sqrt{5}}{GM_im_i^2}=2.618GM_i$. If we set any of the $n_i$s to $3$, then the corresponding $x^{\psi_{n_i}}_\mathrm{max}=6.525GM_i$ will go outside its nearest outer neighbour. If we set all of the $n_i$s to $3$, then the corresponding wave functional of the whole system would have a too fat non-zero tail outside the classic horizon $r_h=2GM_\mathrm{tot}$. The object described by the corresponding wave functional would have matter contents distributed in a spherical spatial region of radius $r=6.525GM_\mathrm{tot}$ instead of being localised inside the classic horizon. So we will not take such highly excited quantum state as the microscopic state of black holes defined by the classic horizon $r_h=2GM_\mathrm{tot}$. We will call the microscopic state defined by the fundamental shell decomposition and pure $2$ radial quantum number as the fundamental state of the classic black hole. For latter usages, we introduce the following diagram to represent this fundamental state
\bea{}
&&\hspace{-7mm}M_{k~}\textcolor{cyan}{\rule[-4pt]{55mm}{4mm}\,\rule[-4pt]{5mm}{4mm}\hspace{-5mm}}m_k,n_k{=}2
\nonumber
\\
&&\hspace{-7mm}M_{\hspace{-1pt}k^{\!-}}\textcolor{cyan}{\rule[-4pt]{50mm}{4mm}\,\rule[-4pt]{5mm}{4mm}\hspace{-5mm}}m_{\hspace{-1pt}k^{\!-}},n_{k^{\!-}}{=}2
\nonumber
\\
&&\hspace{-7mm}M_{i~}\cdots\cdots\rule{20mm}{0pt}\textcolor{cyan}{\rule[-4pt]{5mm}{4mm}\hspace{-5mm}}m_i\,,n_{i}{=}2
\label{basicScheme}
\\
&&\hspace{-7mm}M_{1~}\textcolor{cyan}{\,\rule[-4pt]{5mm}{4mm}\,\rule[-4pt]{5mm}{4mm}\hspace{-5mm}}m_1,n_1{=}2
\nonumber
\\
&&\hspace{-7mm}M_{0~}\textcolor{cyan}{\rule[-4pt]{0mm}{4mm}\,\rule[-4pt]{5mm}{4mm}\hspace{-5mm}}m_{0},n_0{=}2
\nonumber
\eea
where $k\!^{-}{\equiv}k{-}1$ and $GM_im_i=\sqrt{2}$ for all $i$. By this representation, the most general radial excitation scheme can be written as
\bea{}
&&\hspace{-7mm}M_{k~}\textcolor{cyan}{\rule[-4pt]{53mm}{4mm}\textcolor{red}{\xrightarrow{~}}\rule[-4pt]{5mm}{4mm}\hspace{-5mm}}m_k,n_k{=}2
\nonumber
\\
&&\hspace{-7mm}M_{\hspace{-1pt}k^{\!-}}\textcolor{cyan}{\rule[-4pt]{48mm}{4mm}\textcolor{red}{\xrightarrow{~}}\rule[-4pt]{5mm}{4mm}\hspace{-5mm}}m_{\hspace{-1pt}k^{\!-}},\textcolor{red}{n_{k^{\!-}}{\geqslant}2}
\nonumber
\\
&&\hspace{-7mm}M_{i~}\cdots\cdots\rule{18mm}{0pt}\textcolor{red}{\xrightarrow{~~~}}\textcolor{cyan}{\rule[-4pt]{5mm}{4mm}\hspace{-5mm}}m_i\,,\textcolor{red}{n_{i}{\geqslant}2}
\\
&&\hspace{-7mm}M_{1~}\textcolor{cyan}{\,\rule[-4pt]{5mm}{4mm}\,\textcolor{red}{\xrightarrow{~~~~}}\rule[-4pt]{5mm}{4mm}\hspace{-5mm}}m_1,\textcolor{red}{n_1{\geqslant}2}
\nonumber
\\
&&\hspace{-7mm}M_{0~}\textcolor{cyan}{\rule[-4pt]{0mm}{4mm}\,\textcolor{red}{\xrightarrow{~~~~}}\rule[-4pt]{5mm}{4mm}\hspace{-5mm}}m_{0},\textcolor{red}{n_0{\geqslant}2}
\nonumber
\eea
However, by the explicit form of each shell's wave function \eqref{singleShellWvfunction}, it can be easily proved that one cannot find any radial quantum numbers' combination $\{n_i\}$ which is different from that of the fundamental state \eqref{basicScheme} and for all $i$, the position of $|\psi_i^2|$'s maximal value does not go outside that of the shell $m_{i+1}$'s $|\psi^2_{i+1}|$. This means that the fundamental state of the system is unique. 

By choosing two or more inner shells and merging them into some new single shell, we have ways to excite the inner structure of the fundamental state without change the total ADM-mass/energy of the system. The most simple choice is selecting two neighbouring shells and merging them into a single one. For example, merging the outmost two shells and set $n_{k^{-}}{\approx}4$
\bea{}
&&\hspace{-7mm}M_{\hspace{-1pt}k^{\!-}}\textcolor{cyan}{\rule[-4pt]{47mm}{4mm}\,\rule[-4pt]{10mm}{4mm}\hspace{-7mm}}m_{\hspace{-1pt}k^{\!-}}~,n_{k^{\!-}}{\approx}4
\nonumber
\\
&&\hspace{-7mm}M_{i~}\cdots\cdots\rule{20mm}{0pt}\textcolor{red}{\xrightarrow{~~}}\textcolor{cyan}{\rule[-4pt]{5mm}{4mm}\hspace{-5mm}}m_i\,,\textcolor{red}{n_{i}{\geqslant}2}
\\
&&\hspace{-7mm}M_{1~}\textcolor{cyan}{\,\rule[-4pt]{5mm}{4mm}\,\textcolor{red}{\xrightarrow{~~}}\rule[-4pt]{5mm}{4mm}\hspace{-5mm}}m_1,\textcolor{red}{n_1{\geqslant}2}
\nonumber
\\
&&\hspace{-7mm}M_{0~}\textcolor{cyan}{\rule[-4pt]{0mm}{4mm}\,\textcolor{red}{\xrightarrow{~~}}\rule[-4pt]{5mm}{4mm}\hspace{-5mm}}m_{0},\textcolor{red}{n_0{\geqslant}2}
\nonumber
\eea
or more generally
\bea{}
&&\hspace{-7mm}M_{\hspace{-1pt}k^{\!-}}\textcolor{cyan}{\rule[-4pt]{52mm}{4mm}\,\rule[-4pt]{5mm}{4mm}\hspace{-5mm}}m_{\hspace{-1pt}k^{\!-}},n_{k^{\!-}}{=}2
\nonumber
\\
&&\hspace{-7mm}M_{i~}\cdots\cdots\rule{15mm}{0pt}\textcolor{cyan}{\rule[-4pt]{10mm}{4mm}\hspace{-7mm}}m_i~~~,~n_{i}{\approx}4
\\
&&\hspace{-7mm}M_{1~}\textcolor{cyan}{\,\rule[-4pt]{5mm}{4mm}\,\textcolor{red}{\xrightarrow{~~}}\rule[-4pt]{5mm}{4mm}\hspace{-5mm}}m_1,\textcolor{red}{n_2{\geqslant}2}
\nonumber
\\
&&\hspace{-7mm}M_{0~}\textcolor{cyan}{\rule[-4pt]{0mm}{4mm}\,\textcolor{red}{\xrightarrow{~~}}\rule[-4pt]{5mm}{4mm}\hspace{-5mm}}m_{0},\textcolor{red}{n_1{\geqslant}2}
\nonumber
\eea
$n_{k^{\!-}}{\equiv}\frac{GM_{k\!^-\!}(m_{k\!^-\!}{+}m_k)}{(2-2\gamma_{k\!^-\!}^2)^{1/2}\hbar}{\approx}4$ or $n_i{\equiv}\frac{GM_{i}(m_{i}{+}m_{i\!^+\!})}{(2-2\gamma_{i}^2)^{1/2}\hbar}{\approx}4$ means that as the two shells are merged, $n_{k\!^-\!}$ and $n_i$ can be set to $4$ if we tune the value of $\gamma_{k\!^-}$ or $\gamma_i$ to some value which is larger than their value in the fundamental state. This is rational because when we merge two shells, the resultant single shell must be put onto a state whose radial quantum number is roughly the summation of the merger-before members' quantum number to assure the global maximal value of the resultant shell's wave function happens outside the horizon $r^{i'}_h=2GM_{i'}$. Before the two shells' recombination, all $n_{i}$ equals to $2$. The radial quantum number of shells inside the recombined shell can now be set to values greater than 2 because as the latter takes now a larger radial quantum number, the formers have the possibility to be excited to higher radial quantum number level and still keep their global maximal value inside the that of the latter. This means that a given recombined shell decomposition scheme could have more than one radial excitation ways.

The most general shell decomposition and radial excitation scheme can be generated from the fundamental state by shell recombination and radial quantum numbers' reassigning. The resultant microscopic state of the system will have equal total mass as but different inner profiles from the fundamental one. They can be represented as
\bea{}
&&\hspace{-7mm}M_{\ell~}\textcolor{cyan}{\rule[-4pt]{50mm}{4mm}\,\rule[-4pt]{10mm}{4mm}\hspace{-7mm}}m_\ell~~,~
n_\ell=n_\ell^\mathrm{min}
\nonumber
\\
&&\hspace{-7mm}M_{i~}\cdots\cdots\rule{12mm}{0pt}\textcolor{red}{\xrightarrow{~}}\textcolor{cyan}{\rule[-4pt]{12mm}{4mm}\hspace{-8mm}}m_i~~~,~\textcolor{red}{n_{i}{\geqslant}2}
\\
&&\hspace{-7mm}M_{1~}\textcolor{cyan}{\,\rule[-4pt]{5mm}{4mm}\,\textcolor{red}{\xrightarrow{~~}}\rule[-4pt]{5mm}{4mm}\hspace{-5mm}}m_1,\textcolor{red}{n_1{\geqslant}2}
\nonumber
\\
&&\hspace{-7mm}M_{0~}\textcolor{cyan}{\rule[-4pt]{0mm}{4mm}\,\textcolor{red}{\xrightarrow{~~}}\rule[-4pt]{5mm}{4mm}\hspace{-5mm}}m_{0},\textcolor{red}{n_0{\geqslant}2}
\nonumber
\eea
where $n_\ell^\mathrm{min}$ takes the minimal possible value so that $|\psi_\ell^2|$'s global maximal happens outside the horizon $r_h=2GM_\mathrm{tot}$. The number of shell recombination ways from the fundamental state is the number of ordered partition of integer $k=\frac{GM_\mathrm{tot}^2}{2\sqrt{2}}$, which is $2^{k-1}$ exactly, including the fundamental state itself. So the total number of microscopic state of the system can be written as
\beq{}
W=\#\mathrm{{\bf\!p}artition}\{m_i\}\otimes\mathrm{{\bf\!e}xitation}\{n_i\}=2^{k-1}\cdot X(k)
\label{WdefCalculationA}
\eeq
where $X(k)$ is a correction factor which arises from the non-uniqueness of the radial quantum number assigning scheme in the specified shell decomposition scheme. In the case of neglecting the requirement that $|\psi^2_i|$'s global maximal value happens inside that of $|\psi^2_{i+1}|$, references \cite{dfzeng2021} proves that $X(k)$, which is named $F(k)$ there, grows exponentially as $k$ increases. However, when this requirement is considered, it can be verified that $F(k)$'s growing trend is at most polynomial way. See the next section for a concrete working example.

So by the distinguishability definition of \eqref{distinguishabilityRule} and the result \eqref{WdefCalculationA}, we can now claim that to precision $\varepsilon$, the number of all distinguishable ways of decomposing a collapsing star with asymptotic horizon into concentric shells $\{m_i\}$ and setting each shell's radial quantum number $\{n_i\}$ is
 \beq{}
 W{=}\exp\{c(\varepsilon)[\frac{A}{4G}{+}\ln X(A)]\},A{=}4\pi (2GM_\mathrm{tot})^2
 \label{areaLawFormula}
 \eeq
 where $M_\mathrm{tot}$ is only required to be mildly larger than $G^{-\frac{1}{2}}{\equiv}M_\mathrm{pl}$.
Since the decomposition precision parameter $\varepsilon$ linearly affects the number of shells of the fundamental decomposition, $c(\varepsilon)$ is a linear function of $\varepsilon$. This means that we can easily get the logarithmically corrected Bekenstein-Hawking formula with $c(\varepsilon)=1$ by simply setting $\varepsilon\approx\frac{1}{8\sqrt{2}\pi}$. To get the explicit formula of the logarithmic correction $X(A)$, we need quantitatively work out the relation between the number of radial quantum number assigning scheme $\{n_i\}$ and the collapsing star's total mass $M_\mathrm{tot}$ for specified shell decomposition $\{m_i\}$ under the constraint that for all $i$, the global maximal value of $|\psi^2_i|$ happens not outside that of $|\psi^2_{i+1}|$, and the global maximal value of all $|\psi^2_i|$ happens outside the classic horizon of $r^i_h=2GM_i$, with $n_{i-outmost}$ takes the minimal allowed value. This may be a challenging work but operable in principle.

Equation \eqref{enQuantizConditionSchwz} or \eqref{radialQuantizeCondition} implies that the outmost shell has a minimal allowed mass
\bea{}
&&\hspace{-5mm}\frac{GM_\mathrm{tot}m_\mathrm{outmost}}{(2-2\gamma_\mathrm{outmost}^2)^{1/2}}=n_\mathrm{outmost}^\mathrm{min}=2\Rightarrow
\\
&&\hspace{-5mm}m_\mathrm{outmost}=\frac{2(2-2\gamma^2_\mathrm{outmost})^\frac{1}{2}}{GM_\mathrm{tot}}
\eea
Since the shell with $\gamma_\mathrm{outmost}{=}0$ has the maximal probability be found on the horizon and that with $\gamma_\mathrm{outmost}=1$ has the maximal probability be found on the infinity $r$ region. Exact horizon boundary condition implies that $\gamma_\mathrm{outmost}{=}0$. This means that
\beq{}
m^\mathrm{min}_\mathrm{outmost}=\frac{2\sqrt{2}}{GM_\mathrm{tot}}
\label{mMinOutmost}
\eeq
This is also the typical mass of most of the outer layered shells. This explains why we have chances to get the area law entropy by counting the bulk motion degrees of freedom of the black hole matter contents. Because the number of such shells reads
\beq{}
N\sim\frac{M_\mathrm{tot}}{(GM_\mathrm{tot})^{-1}}\propto GM^2
\eeq
Note that for all macroscopic black holes, the value of $m^\mathrm{min}_\mathrm{outmost}$ is far less than the standard model particles which consists of the collapsing star. So what the Bekenstein-Hawking entropy reflects is not the motion degrees of freedom of these particles, but the collection motion modes of them which we call as composite shells. This fact is pointed out and emphasised in our series of works \cite{dfzeng2017,dfzeng2018a,dfzeng2018b,dfzeng2020,dfzeng2021,dfzeng2022}.

Equation \eqref{mMinOutmost} also tells us that $m^\mathrm{min}_\mathrm{outmost}$ has the same order as the hawking temperature of a mass $M_\mathrm{tot}$ black hole. This means that this shell can be easily radiated away as hawking radiations. FIG.\ref{figWFsground} displays the wave function modular square of the fundamental microscopic state of a $3M_\mathrm{pl}$ black hole caused by a dust ball's gravitational collapse. From the figure we easily see the out most shell has very large probability be found outside the horizon. Of course this outmost shell's appearance outside the horizon itself is not hawking radiation. The true hawking radiation arises from the microscopic state's change of the system and can be described as the spontaneous radiation of usual atoms with explicitly hermitian hamiltonian, which is named gravitation induced spontaneous radiation in \cite{dfzeng2022},  see references \cite{dfzeng2018a,dfzeng2018b,dfzeng2021} for more earlier and detailed investigations.

The physical picture and calculations up to here provide a quantum mechanics based general relativity interpretation for the Bekenstein-Hawking entropy and area law formula. This interpretation contrasts with the common belief that Bekenstein-Hawking entropy has pure geometric origin, i.e. the fundamental degrees of freedom of black holes is carried by their horizon, each planck unit of area element carries one bit of information. However, our calculation shows that such fundamental degrees of freedom may completely be carried by the bulk motion modes of the matter contents. As long as we find the right carrier the area law formula flows logically and naturally. Since in a collapsing star consists of dust only, no other forces except gravitation is involved, the matter's motion and fluctuation is an exact and faithful reflection of the geometric feature of the spacetime underground. As results, the quantisation of these matters' motion is also the quantisation of spacetime itself. In a sense what we provide in this work is a quantum gravitation theory indeed. Since it focuses on the origin of Bekenstein Hawking entropy exclusively, we consider it a mini-version of some underlying full quantum gravitation theories. 

\section{A working example for the quantisation}
\label{exampleQuantisation}

Let us consider the inner structure's quantisation of an $M_\mathrm{tot}=3M_\mathrm{pl}$ black hole as a concrete example for the microscopic state definition ideas above. For this black hole, to precision $\varepsilon=1$, its matter contents can be divided into 4 distinguishable concentric shells at most, under the condition that $n_i{=}2$, $\gamma_i{=}0$ for all $i$. The fundamental state of the whole system is defined by the following shell mass assigning and radial quantum level number's setting scheme, 
\bea{}
&&\hspace{-5mm}\{m_i\}=\{1.25118, 0.718127, 0.559288, 0.471405\}
\\
&&\hspace{-5mm}\{M_i\}=\{1.25118, 1.96931, 2.5286, 3\}
\\
&&\hspace{-5mm}\{\frac{GM_im_i}{\sqrt{2}}\}=\{1, 1, 1,1\}
\\
&&\hspace{-5mm}\{(1-\gamma_i^2)^\frac{1}{2}\}=\{0.5, 0.5, 0.5,0.5\}
\\
&&\hspace{-5mm}\{n_i\}=\{2, 2, 2,2\}
\\
&&\hspace{-5mm}\{\psi_i\}\propto\{e^{-\bar{x}}\bar{x}\mathrm{LaguerreL}_0^1(2\bar{x})\}
\label{waveFunctionGroundstate}
\\
&&\hspace{-5mm}\{x_\mathrm{max}^{\psi_i}\}=\{2.67, 5.16, 6.62, 7.85\}
\eea
Note that in the wave function of \eqref{waveFunctionGroundstate}, $\bar{x}{\equiv}m_i(2{-}2\gamma_i^2)^\frac{1}{2}x$.
By the representation method of \eqref{basicScheme}, this ground state can be represented as
\bea{}
&&\hspace{-7mm}M_{3~}\textcolor{cyan}{\rule[-4pt]{32mm}{4mm}\,\rule[-4pt]{5mm}{4mm}\hspace{-5mm}}m_3,n_3{=}2
\\
&&\hspace{-7mm}M_{2~}\textcolor{cyan}{\rule[-4pt]{24mm}{4mm}\,\rule[-4pt]{7mm}{4mm}\hspace{-5mm}}m_{\hspace{-1pt}2},n_{2}{=}2
\nonumber
\\
&&\hspace{-7mm}M_{1~}\textcolor{cyan}{\rule[-4pt]{13mm}{4mm}\,\rule[-4pt]{10mm}{4mm}\hspace{-5mm}}m_1,n_1{=}2
\nonumber
\\
&&\hspace{-7mm}M_{0~}\textcolor{cyan}{\rule[-4pt]{13mm}{4mm}\hspace{-5mm}}m_{0},n_0{=}2
\nonumber
\eea

\begin{figure}[h]\begin{center}
\includegraphics[totalheight=50mm]{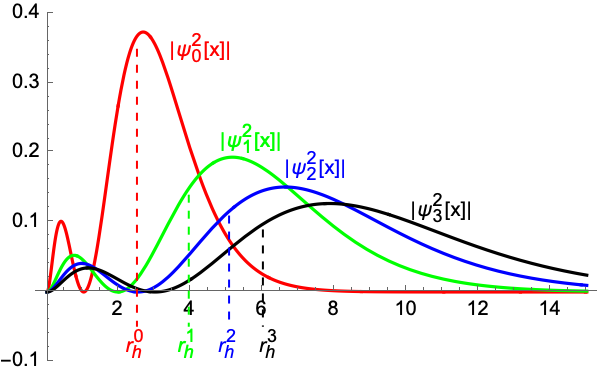}
\end{center}
\caption{The modular square of the ground state wave function of an $M=3M_\mathrm{pl}$ black hole.}
\label{figWFsground}
\end{figure}
FIG.\ref{figWFsground} displays the modular square of the wave function of each sub-shell. In the figure, three things are worthy of noticing. The first is, the positions of the global maximal value of each sub-shell's wave function are ordered from inside to the outside. The second is, the global maximal value of all shell $m_i$'s wave function happens outside $r^i_h$. This is nothing but the requirement that classically each shell lies outside the horizon  $r^i_h=2G\sum_{i'=0}^im_{i'}$ determined by the matter mass they wrap. The third is, the probability of  each shells' being found outside the global horizon defined by $r_h=2GM_\mathrm{tot}$ is rather remarkable. Keeping the condition $x^{\psi_i}_\mathrm{max}<x^{\psi_{i+1}}_\mathrm{max},\forall i$ and exciting all shells $m_i$ to their higher level quantum state uniformly by setting $n_i=3$ or more larger is possible, but that would make the probability of these shell's being found outside the global horizon becomes too larger than the ground state characterised by $n_i=2,\forall i$. 

Now if we let some but not all of the inner shells such as $m_0$ to be excited, we will find that even only let $n_0=3$, the global maximal value of $\psi_0(x)$ would happen outside that of $\psi_1(x)$. So under the requirement that all other shell's quantisation conditions are kept invariant, i.e. $\frac{GM_im_i}{\sqrt{2}}=2$($i{\neq}0\&1$), the only possible way of exciting $m_0$ is letting it be bounded with $m_1$ as a single shell. Similarly, if we want excite other one or more shells independently, the best way is also to let them be bounded with their closest outer neighbour into a single shell. For the $M=3M_\mathrm{pl}$ black hole whose fundamental state is defined by the four $n_i=2$($i{=}0,1,2,3$) shells, we have only $7=2^{4-1}-1$ possible ways to recombine those shells. Let us consider them one by one.

Recombination A, $\{m_0+m_1{\equiv}m'_0,m_2{\equiv}m'_1,m_3{\equiv}m_2'\}$, remember in the wave function \eqref{waveFunctionRecA}, $\bar{x}{\equiv}m_i(2{-}2\gamma_i^2)^\frac{1}{2}x$. This is the same as \eqref{waveFunctionGroundstate} and similarly in the other recombination schemes.
\bea{}
&&\hspace{-5mm}\{m'_i\}=\{1.96931,0.559288,0.471405\}
\\
&&\hspace{-5mm}\{M'_i\}=\{1.96931,2.5286,3.\}
\\
&&\hspace{-5mm}\{\frac{GM'_im'_i}{\sqrt{2}}\}=\{2.74228,1,1\}
\eea
\bea{}
&&\hspace{-5mm}\{(1-{\gamma'}_i^2)^\frac{1}{2}\}=\{\Gamma_0, 0.5, 0.5\},0\leqslant\Gamma_0\leqslant1
\\
&&\hspace{-5mm}\{n'_i\}=\{2.74228/\Gamma_0, 2,2\}
\\
&&\hspace{-5mm}\{\psi'_i\}\propto\{e^{-\bar{x}}\bar{x}\mathrm{LaguerreL}_{n_i-1}^1(2\bar{x})\}
\label{waveFunctionRecA}
\\
&&\hspace{-5mm}\{x_\mathrm{max}^{\psi_i'}\}=\{3.94{\leqslant}x^{\psi'_0}_\mathrm{max}{\leqslant}6.62, 6.62, 7.85\}
\eea
To satisfy the constraint $2GM'_0{=}3.94{\leqslant}x^{\psi'_0}_\mathrm{max}{\leqslant}6.62$, we find that $n'_0=5$ are allowed, other values are not. So this shell recombination scheme contributes only one microscopic state to the system. By the representation method of \eqref{basicScheme}, this state can be represented as
\bea{}
&&\hspace{-7mm}M'_{2~}\textcolor{cyan}{\rule[-4pt]{32mm}{4mm}\,\rule[-4pt]{5mm}{4mm}\hspace{-5mm}}m'_2,n_2{=}2
\\
&&\hspace{-7mm}M'_{1~}\textcolor{cyan}{\rule[-4pt]{24mm}{4mm}\,\rule[-4pt]{7mm}{4mm}\hspace{-5mm}}m'_1,n_{1}{=}2
\nonumber
\\
&&\hspace{-7mm}M'_{0~}\textcolor{red}{\xrightarrow{~~}}\textcolor{cyan}{\rule[-4pt]{19mm}{4mm}\hspace{-5mm}}m'_0,\textcolor{red}{n_0{=}5}
\nonumber
\eea

Recombination B, $\{m_0{\equiv}m'_0,m_1{+}m_2{\equiv}m'_1,m_3{\equiv}m_2'\}$, still remember that $\bar{x}{\equiv}m_i(2{-}2\gamma_i^2)^\frac{1}{2}x$
\bea{}
&&\hspace{-5mm}\{m'_i\}=\{1.25118,1.27742,0.471405\}
\\
&&\hspace{-5mm}\{M'_i\}=\{1.25118,2.5286,3\}
\\
&&\hspace{-5mm}\{\frac{GM'_im'_i}{\sqrt{2}}\}=\{1.10694,2.284,1\}
\eea
\bea{}
&&\hspace{-5mm}\{(1-{\gamma'}_i^2)^\frac{1}{2}\}=\{0.55347,\Gamma_1, 0.5\},0\leqslant\Gamma_1\leqslant1
\\
&&\hspace{-5mm}\{n'_i\}=\{2,2.28406/\Gamma_1,2\}
\\
&&\hspace{-5mm}\{\psi'_i\}\propto\{e^{-\bar{x}}\bar{x}\mathrm{LaguerreL}_{n_i{-}1}^1(2\bar{x})\}
\\
&&\hspace{-5mm}\{x_\mathrm{max}^{\psi_i'}\}=\{2.67, 5.06{\leqslant}x^{\psi'_1}_\mathrm{max}{\leqslant}7.85, 7.85\}
\eea
To satisfy the condition $2GM'_2{=}5.06<x^{\psi'_1}_\mathrm{max}<7.85$, we find that $n'_1=4$ are allowed, other values are not. This means that the shell recombination scheme B also contributes only one microscopic state to the system. The diagram representation of this state can be plotted as follows
\bea{}
&&\hspace{-7mm}M'_{2~}\textcolor{cyan}{\rule[-4pt]{32mm}{4mm}\,\rule[-4pt]{5mm}{4mm}\hspace{-5mm}}m'_2,n_2{=}2
\\
&&\hspace{-7mm}M'_{1~}\textcolor{cyan}{\rule[-4pt]{13mm}{4mm}\textcolor{red}{\xrightarrow{~~}}\rule[-4pt]{14mm}{4mm}\hspace{-5mm}}m'_1,\textcolor{red}{n_{1}{=}4}
\nonumber
\\
&&\hspace{-7mm}M'_{0~}\textcolor{cyan}{\rule[-4pt]{13mm}{4mm}\hspace{-5mm}}m'_0,n_0{=}2
\nonumber
\eea

Recombination C, $\{m_0{\equiv}m'_0,m_1{\equiv}m'_1,m_2{+}m_3{\equiv}m_2'\}$
\bea{}
&&\hspace{-5mm}\{m'_i\}=\{1.25118,0.718127,1.03069\}
\\
&&\hspace{-5mm}\{M'_i\}=\{1.25118,1.96931,3\}
\\
&&\hspace{-5mm}\{\frac{GM'_im'_i}{\sqrt{2}}\}=\{1.10694,1,2.18643\}
\eea
\bea{}
&&\hspace{-5mm}\{(1-{\gamma'}_i^2)^\frac{1}{2}\}=\{0.55347, 0.5,\Gamma_2\},0\leqslant\Gamma_2\leqslant1
\\
&&\hspace{-5mm}\{n'_i\}=\{2,2,2.18643/\Gamma_2\}
\\
&&\hspace{-5mm}\{\psi'_i\}\propto\{e^{-\bar{x}}\bar{x}\mathrm{LaguerreL}_{n_i-1}^1(2\bar{x})\}
\\
&&\hspace{-5mm}\{x_\mathrm{max}^{\psi_i'}\}=\{2.67, 5.16, 6{\leqslant}x^{\psi'_2}_\mathrm{max}\}
\eea
To satisfy the condition $2GM'_2{=}6{\leqslant}x^{\psi'_2}_\mathrm{max}$, we find that $n'_2=4$ is the lowest requirement. Other greater than $4$ values for $n'_2$ could also satisfy this constraint. But the corresponding wave function will have a very fat non-zero tail outside the global horizon thus contradict with definition of black hole in general relativity. We will exclude such state as the microscopic state of the black holes characterised by the classic horizon $r_h=2GM_\mathrm{tot}$. This means that the shell recombination scheme C also contributes only one microscopic state to the system. The diagram representation of this state can be plotted as follows
\bea{}
&&\hspace{-7mm}M'_{2~}\textcolor{cyan}{\rule[-4pt]{25mm}{4mm}\textcolor{red}{\xrightarrow{~~}}\rule[-4pt]{9mm}{4mm}\hspace{-5mm}}m'_2,\textcolor{red}{n_2{=}4}
\\
&&\hspace{-7mm}M'_{1~}\textcolor{cyan}{\rule[-4pt]{13mm}{4mm}\,\rule[-4pt]{11mm}{4mm}\hspace{-5mm}}m'_1,n_{1}{=}2
\nonumber
\\
&&\hspace{-7mm}M'_{0~}\textcolor{cyan}{\rule[-4pt]{13mm}{4mm}\hspace{-5mm}}m'_0,n_0{=}2
\nonumber
\eea

Recombination D, $\{m_0{+}m_1{+}m_2{\equiv}m'_0,m_3{\equiv}m'_1\}$
\bea{}
&&\hspace{-5mm}\{m'_i\}=\{2.5286,0.4714\}
\\
&&\hspace{-5mm}\{M'_i\}=\{2.5286,3\}
\\
&&\hspace{-5mm}\{\frac{GM'_im'_i}{\sqrt{2}}\}=\{4.5211,1\}
\eea
\bea{}
&&\hspace{-5mm}\{(1-{\gamma'}_i^2)^\frac{1}{2}\}=\{\Gamma_0, 0.5\},0\leqslant\Gamma_0\leqslant1
\\
&&\hspace{-5mm}\{n'_i\}=\{4.5211/\Gamma_0,2\}
\\
&&\hspace{-5mm}\{\psi'_i\}\propto\{e^{-\bar{x}}\bar{x}\mathrm{LaguerreL}_{n_i-1}^1(2\bar{x})\}
\\
&&\hspace{-5mm}\{x_\mathrm{max}^{\psi_i'}\}=\{5.06{\leqslant}x^{\psi'_0}_\mathrm{max}{\leqslant}7.85,7.85\}
\eea
To satisfy the condition $2GM'_0{=}5.06{\leqslant}x^{\psi'_0}_\mathrm{max}{\leqslant}7.85$, we find that $n'_0=7,8$ are both allowed. Other values are not. So this recombination contributes 2 microscopic states to the system.  The diagram representation of these two states can be plotted as follows
\bea{}
&&\hspace{-7mm}M'_{1~}\textcolor{cyan}{\rule[-4pt]{25mm}{4mm}\,\rule[-4pt]{5mm}{4mm}\hspace{-5mm}}m'_1,n_1{=}2
\\
&&\hspace{-7mm}M'_{0~}\textcolor{red}{\xrightarrow{~~~}}\textcolor{cyan}{\rule[-4pt]{19mm}{4mm}\,\hspace{-5mm}}m'_0,\textcolor{red}{n_0{=}7,8}
\nonumber
\eea

Recombination E, $\{m_0{+}m_1{\equiv}m'_0,m_2{+}m_3{\equiv}m'_1\}$
\bea{}
&&\hspace{-5mm}\{m'_i\}=\{1.96931, 1.03069\}
\\
&&\hspace{-5mm}\{M'_i\}=\{1.96931, 3\}
\\
&&\hspace{-5mm}\{\frac{GM'_im'_i}{\sqrt{2}}\}=\{2.742,2.186\}
\eea
\bea{}
&&\hspace{-5mm}\{(1-{\gamma'}_i^2)^\frac{1}{2}\}=\{\Gamma_0, \Gamma_1\},0\leqslant\Gamma_0,\Gamma_1\leqslant1
\\
&&\hspace{-5mm}\{n'_i\}=\{2.742/\Gamma_0,2.186/\Gamma_1\}
\\
&&\hspace{-5mm}\{\psi'_i\}\propto\{e^{-\bar{x}}\bar{x}\mathrm{LaguerreL}_{n_i-1}^1(2\bar{x})\}
\\
&&\hspace{-5mm}\{x_\mathrm{max}^{\psi_i'}\}=\{3.94{\leqslant}x^{\psi'_0}_\mathrm{max},6{\leqslant}x^{\psi'_1}_\mathrm{max}\}
\eea
To satisfy the constraints $2GM'_0{=}3.94{\leqslant}x^{\psi'_0}_\mathrm{max}$, $2GM'_1{=}$ $6{\leqslant}x^{\psi'_1}_\mathrm{max}$ and $x^{\psi'_0}_\mathrm{max}<x^{\psi'_1}_\mathrm{max}$, we find that $n'_0=5$, $n'_1=4$ are allowed. Other values for $n'_0$ and $n'_1$ which satisfy this three constraints are also possible, but all of them have the problem of giving the corresponding wave function a too fat nonzero tail outside the horizon thus contradict with the definition of black holes characterised by their classic horizon. So such highly excited state will not be considered as the microscopic state of the class black holes. The diagram representation of this state can be plotted as follows
\bea{}
&&\hspace{-7mm}M'_{1~}\textcolor{cyan}{\rule[-4pt]{20mm}{4mm}\textcolor{red}{\xrightarrow{~~}}\rule[-4pt]{8mm}{4mm}\hspace{-5mm}}m'_1,\textcolor{red}{n_1{=}4}
\\
&&\hspace{-7mm}M'_{0~}\textcolor{red}{\xrightarrow{~~~}}\textcolor{cyan}{\rule[-4pt]{14mm}{4mm}\,\hspace{-5mm}}m'_0,\textcolor{red}{n_0{=}6}
\nonumber
\eea

Recombination F, $\{m_0{\equiv}m'_0, m_1{+}m_2{+}m_3{\equiv}m'_1\}$
\bea{}
&&\hspace{-5mm}\{m'_i\}=\{1.25118, 1.74882\}
\\
&&\hspace{-5mm}\{M'_i\}=\{1.25118, 3\}
\\
&&\hspace{-5mm}\{\frac{GM'_im'_i}{\sqrt{2}}\}=\{1.10694,3.70981\}
\eea
\bea{}
&&\hspace{-5mm}\{(1-{\gamma'}_i^2)^\frac{1}{2}\}=\{0.55347, \Gamma_1\},0\leqslant\Gamma_1\leqslant1
\\
&&\hspace{-5mm}\{n'_i\}=\{2, 3.70981/\Gamma_1\}
\\
&&\hspace{-5mm}\{\psi'_i\}\propto\{e^{-\bar{x}}\bar{x}\mathrm{LaguerreL}_{n_i-1}^1(2\bar{x})\}
\\
&&\hspace{-5mm}\{x_\mathrm{max}^{\psi_i'}\}=\{2.67, 6{\leqslant}x^{\psi'_1}_\mathrm{max}\}
\eea
To satisfy the condition $2GM'_1{=}6{\leqslant}x^{\psi'_1}_\mathrm{max}$, we find that $n'_1=6$ is the lowest level required. Other greater than $6$ levels for $n'_1$ which satisfy this condition are also possible, but has the problem of giving the corresponding wave function a too fat tail outside the classic horizon. So they will still not be taken as the microscopic state of class black holes. The diagram representation of this state can be plotted as follows
\bea{}
&&\hspace{-7mm}M'_{1~}\textcolor{cyan}{\rule[-4pt]{12mm}{4mm}\textcolor{red}{\xrightarrow{~~}}\rule[-4pt]{16mm}{4mm}\hspace{-5mm}}m'_1,\textcolor{red}{n_1{=}6}
\\
&&\hspace{-7mm}M'_{0~}\textcolor{cyan}{\rule[-4pt]{12mm}{4mm}\,\hspace{-5mm}}m'_0,n_0{=}2
\nonumber
\eea

Recombination G, $\{m_0{+}m_1{+}m_2{+}m_3{\equiv}m'_0\}$
\bea{}
&&\hspace{-5mm}\{m'_i\}=\{3\}, \{M'_i\}=\{3\}
\\
&&\hspace{-5mm}\{\frac{GM'_im'_i}{\sqrt{2}}\}=\{6.36396\}
\eea
\bea{}
&&\hspace{-5mm}\{(1-{\gamma'}_i^2)^\frac{1}{2}\}=\{\Gamma_0\},0\leqslant\Gamma_1\leqslant1
\\
&&\hspace{-5mm}\{n'_i\}=\{6.36396/\Gamma_0\}
\\
&&\hspace{-5mm}\{\psi'_i\}\propto\{e^{-\bar{x}}\bar{x}\mathrm{LaguerreL}_{n_i-1}^1(2\bar{x})\}
\\
&&\hspace{-5mm}\{x_\mathrm{max}^{\psi_i'}\}=\{6{\leqslant}x^{\psi'_0}_\mathrm{max}\}
\eea
To satisfy the condition $2GM'_0{=}6{\leqslant}x^{\psi'_1}_\mathrm{max}$, we find that $n'_0=10$ is the lowest level required. Other greater than $10$ levels for $n'_0$ satisfying this condition are also possible, but has the problem of giving the corresponding wave function a too fat nonzero tail outside the classic horizon. So still can not be taken as microscopic states of the classic black holes. The diagram representation of this state can be plotted as follows
\beq{}
M'_{0~}\textcolor{red}{\xrightarrow{~~}}\textcolor{cyan}{\rule[-4pt]{20mm}{4mm}}\hspace{-5mm}m'_0,\textcolor{red}{n_0{=}10}
\eeq

From the calculations above, we can see that to precision $\varepsilon=1$, an $M=3M_\mathrm{pl}$ black hole can be divided into 4 distinguishable concentric shells at most under the condition that $n_i{=}2$ and $\gamma_i{=}0$ for all $i$. Letting all this four shells lie on the quantum state of $n_i=2$ yields the fundamental state of the system. In this fundamental state, the global maximal value of each shells' wave function modular square happens outside the classic horizon determined by the masses the shell wraps $r^i_h{=}2GM_i$, $M_i{=}\sum_{i'=0}^im_i$. To excite each shell independently and preserve the ordering of the excited shells is equivalent to recombine  shells from the fundamental state and tuning the resultant shell's radial quantum number. Including the fundamental state itself, we have $2^{4-1}=8$ ways to to such a recombination. Except the recombination scheme D, the number of ways to excite all shells in all the recombinations is unique. Generalising to the large mass case, this is our conclusion proved analytically in the previous section, the matter contents of a spherical symmetric black hole of mass $M$(mildly larger than $M_\mathrm{pl}$) with asymptotic horizon $r=2GM$ can be divided into $k=\frac{GM^2}{2\sqrt{2}}$ distinguishable concentric shells to precision $\varepsilon=1$. These fundamental shells can be recombined and excited at $2^{k-1}X(k)$ number of ways. $X(k)$ is at most a polynomial function of $k$.

\section{Black hole complementarity principle}
\label{secComplementarity}

By our proof and calculations above, one can easily understand that in the co-moving observer's description, the gravitational collapse passes through the horizon and Schwarzschild singularity fluently in finite and seeable proper time. The resulting black hole owns an quasi-periodically oscillatory matter core and multiply possible mass profiles which is ergodically accessed by the evolution of system. While in the outside fixed position observers' time definition domain, a collapsing star can infinitely approach but will never contract into the horizon defined by its total mass; the system forms only an asymptotically implementable horizon and asymptotically linear mass function, the microscopic state of the system is characterised by its mass function's deviation from the exact linear profile $M[r]=\frac{r}{2G}$. The gap between this two observers' description is filled up as long as we note that, the horizon and singularity encountered by the co-moving observer happen only in the Future of infinitely Far away Future (FiFF) by the outside fixed position observers' physical time definition. They are realities only in the parallel universe or ensemble spacetime of these outside fixed position observer or detectors.

\begin{figure}[ht]
\includegraphics[totalheight=58mm]{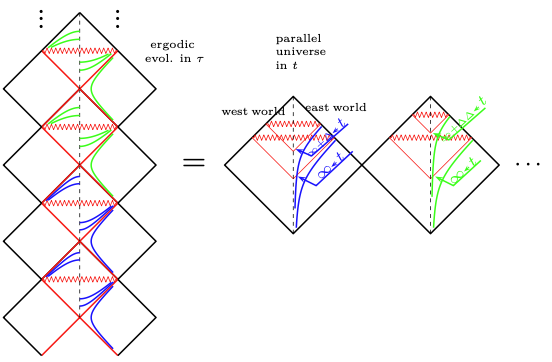}
\caption{Equivalence between two types of observers microscopic state definition for the black holes. Relative to FIG.\ref{figPCfxoInnStructABC}, the right hand part of this figure adopts the extended rule of conventional Penrose Carter diagram so that the east world and west world of the spacetime are simultaneously displayed in the $\{\tau,\rho\}$ plane.
}\label{figPCcomplementarity}
\end{figure}

This fact naturally brings us to the idea of black hole complementarity principle \cite{complementarity1990thooft,complementarity1993verlinde,complementarity1993schoutens,complementarity1993,complementarity1994}, which conjectures that both the co-moving observer and the outside fixed position observer's description of a collapsing star's evolution are equally true. What we want to complement here is that, when the ergodicity feature of the co-moving observer and its observables' interaction dynamics and the uncertainty principle implied parallel universe or ensemble spacetime interpretation for the outside fixed position observers' measurements are considered, the two types of observer's definition for the black hole microscopic state will not only be equally true, but also be equally complete. In another word, the two observers' description are equivalent. By the extended Penrose-Carter diagram, we display this equivalence in FIG.\ref{figPCcomplementarity}.

In the quantisation stratagem of references \cite{dfzeng2018a,dfzeng2018b,dfzeng2020,dfzeng2021} we assumed $\gamma_i^2<0$ for each single shell when do the microscopic state number counting for black holes caused by gravitational collapse. Technically, this stratagem is nothing but the requirement that classically all shells consisting of the collapsing star oscillate inside the horizon defined by $r<r^i_h\equiv2GM_i$ so that for these shells both $\dot{t}$ and $\dot{x}$ take imaginary values but the four velocity normalisation condition $h\dot{t}^2-h^{-1}\dot{r}^2=1$ holds just as the shell is moving outside the horizon $r^i_h$. While in our proofs of this work, section \ref{secQuantisation}, this stratatem is given up and the same area law is obtained successfully. Physically this means that we do not need any shell $m_i$ to move inside its horizon $r^i_h=2GM_i$, we only need to consider their motion of falling towards there. The fact that the two proofs lead to the same area law entropy formula implies that, there is an equivalence between the two pictures of (i) the motion of matters consisting of a black hole is their oscillation over-crossing the central point and (ii) the motion of matters consisting of a black hole is its composite shells' falling towards their own's horizon. In the latter picture, the quantum fluctuation implied by the uncertainty principle has the power of causing various shells to escape away from their asymptotical horizon thus making their inside horizon space looks like something just be dug out from the whole spacetime and anti-podal point identified \cite{tHooft2021,tHooft2019,tHooft1996,tHooftISSP2016,tHooft1601,tHooft1605,tHooft1612,tHooft1804,tHooft1809,tHooftFP2018,tHooftISSP2018}.

Although our proof or illustration of the equivalence relation between the two observers' definition of the black hole microscopic states is based on the physical black holes with non-singular inner mass distribution. We emphasise here that, even for the mathematical black holes defined as the exact solutions to the vacuum Einstein equation thus characterised by the apriori horizon and singularity. Similar complementarity principle must also be possible be understood from the two types observers' microscopic state definition. Because as long as such microscopic state are physical, they must be observable in principle. For such black holes, the idea of black hole complementarity principle is challenged by the firewall paradoxes in recent years \cite{fireworksAMPS2012,fireworksAMPS2013}

\begin{figure}[ht]
\includegraphics[totalheight=70mm]{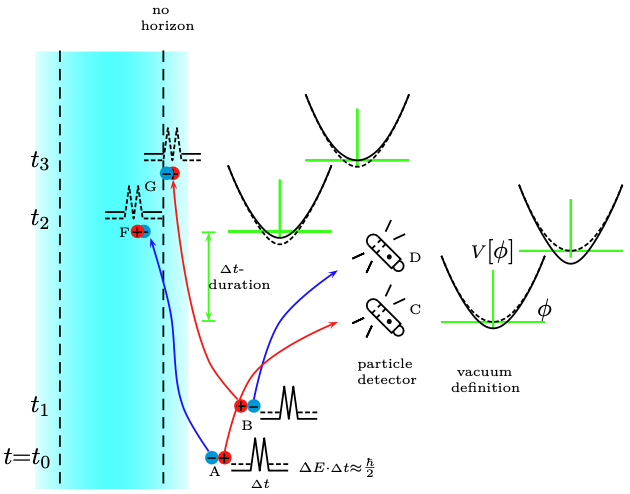}
\caption{The vacuum fluctuation and partial escaping mechanism for hawking radiation is a policy of deficit financing. The cost of the outside Gieger counter's recording hawking particles is the vacuum definition standard's lowering, an analog of the currency releasing standard's lowering in conventional financial systems. An event of particle pair's annihilation would uplift this vacuum standard, analog of the currency releasing standard's up-tuning. The potential curve $V[\phi]$'s dash to solid style change denotes the vacuum definition standard's change.  For a black hole caused by gravitational collapse, the horizon is only an asymptotically realisable but not a truly implemented physic surface so matter particles of the system can be found outside of it. The time separation ${\Delta}t$ between the Gieger counter's recording particle and the ingoing particle's annihilation with matter particles consisting of the collapsing star is very short so that ${\Delta}t{\cdot}{\Delta}E\approx\frac{\hbar}{2}$ is holden very well. }\label{figHwkMechanismCstar}
\end{figure}

By definition, the black hole in the fire wall paradox is nothing but a horizon wrapped singularity and between the horizon and the singularity is a vacuum region with wrongly signatured space-time metric. By the fire wall paradox's reasoning, the hawking particle arises from the particle pair production and partial escaping from the vacuum fluctuation around the horizon. The escapers have nothing to do the microscopic state of the black hole. At late times any given escaper is entangled with both its partner falling into the horizon and the early escaping successful colleagues as a whole. This two side entanglement would break the monodromy of the quantum entanglement thus calls for protections from mechanisms such as firewalls. Obviously, for black holes caused by gravitational collapse, exact horizon and singularity never form to the outside observer or detectors. To them the black holes are inner-structured and characterised by special radial mass profiles. Referring FIG.\ref{figHwkMechanismCstar}, late time hawking particles can be recorded only when the in-going member of the early produced pairs get annihilated with matter particles consisting of the collapsing star because otherwise we will have vacuum decay inevitably. Effectively, this makes each hawking particle looks like arising from the microscopic state change of the black hole instead of completely random fluctuation of the vacuums around it. References \cite{dfzeng2018a,dfzeng2018b,dfzeng2021} provide an explicitly hermitian hamiltonian description for this radiation mechanism which is named gravity induced spontaneous radiation in \cite{dfzeng2022}.

\begin{figure}[ht]
\includegraphics[totalheight=70mm]{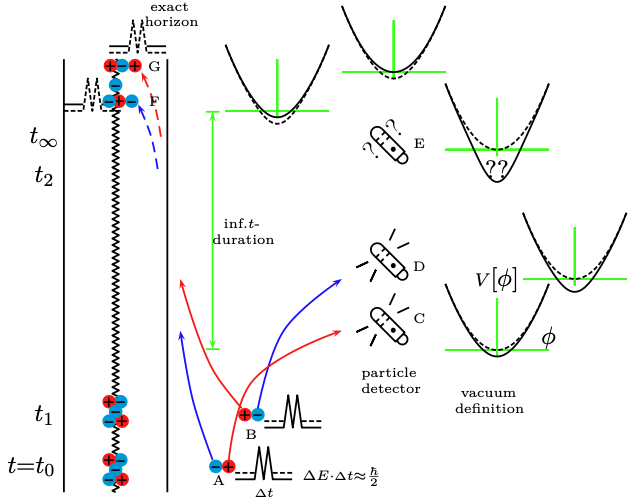}
\caption{For black holes defined as exact solutions to the vacuum Einstein equation, the horizon is absolute and apriori. The in-going particles arising from vacuum pair production needs infinite $t$-time to go across the horizon and annihilate with matters inside to uplift the standard of vacuum definition. During this period, if the outside fixed position Gieger counter records hawking particles continuously, the vacuum definition standard would be observed to be lowered remarkably. This would cause vacuum decay. Particles would be observable everywhere, just like currency inflations in the conventional financial system. So the deficit financing policy interpretion for hawking radiation's production in these black holes does not work at all. That is, the outside fixed position observer could not see the black hole's evaporation as the result of vacuum fluctuation caused particle production and partial escaping.}
\label{figHwkMechanismBH}
\end{figure}

For the black holes defined as exact solutions to the vacuum Einstein equation so carry absolute horizon and singularity apriori, hawking particles cannot be produced continuously as the results of particle pair production and partial escaping from the vacuum fluctuation around the horizon. This is because the hawking radiation recorded by the outside fixed position observers comes from the vacuum definition standard's lowering. The particle pairs annihilation take the responsibility of uplifting this back thus retrieving such standard. To those observers, each member of the particle pairs falling towards the horizon needs infinite time to get across the horizon and reach the central point which accumulates all matter contents of the black hole to annihilate with particles there so that the vacuum definition standard can be uplifted back, see FIG.\ref{figHwkMechanismBH} for intuitions. If an outside fixed position observer records hawking particles continuously in his/her detector, then his/her vacuum definition standard would be lowered continuously without compensation. As results, he/she would see infinite number of particles everywhere as the results of vacuum decaying. So, for black holes defined as exact solutions to the vacuum Einstein equation, the situation imagined in the firewall paradoxes where the black hole radiated half of its mass will not appear at all. In fact the fraction of this black hole's mass change due to hawking radiation is strictly constrained by the uncertainty principle ${\Delta}M{\cdot}{\Delta}t\approx\frac{\hbar}{2}$. 

So the complementarity principle is right enough to be taken as a basis for understanding the black hole inner structures and their microscopic state definition. What the firewall paradox challenges is not the complementarity principle, but the underlying mechanism of hawking particle's generation mechanism. The right mechanisms such as gravity induced spontaneous radiation must contain some ingredients which couples the hawking particle's observational feature such as interaction amplitude with the microscopic state change of the radiating black holes, so that the information encoded in the initial state of the black hole can be transferred into observational features of the hawking particles. See references \cite{dfzeng2018a,dfzeng2018b,dfzeng2021,dfzeng2022} for more concrete and detailed implementation of this idea. 

\section{The physical reality's observer dependence}
\label{secPhysReality}

Why is the singularity theorem chosen by so many people as the basis for understanding the inner structure and microscopic state of black holes in general relativity? Why is it not the complementarity principle interpretation? Pushing aside the establishing time reason, the other one important reason for this status may be people's mistaking of the global viewpoint as local viewpoints and negligence of the physical reality's observer dependence, under the name of the physic law's general coordinate invariance.

It is well known that in a black hole formed through gravitational collapse, the outside fixed position observers will not see the formation of horizon and singularity in their any finite future. These observers can be any outside test particle or probes whose motion is affected by the gravitational field of the to be black hole collapsing star, such as the other partner in a binary merger event and the light rays skimming over the surface of a single black hole to be imaged. During the whole lifetime of these test particle's existence, their motion is affected by the gravitational field provided by the matter contents of the collapsing star being measured. So they see no black holes defined by the horizon and singularity at all, they see just a to be black hole collapsing star.

Due to the co-moving observers existence and the singularity theorem's being proven, many people believe that the fate of all heavy enough collapsing stars in the nature is the horizon wrapped singularity definitely. The logic behind this belief is, the physics laws are invariant under the general coordinate transformation. In that the singularity and horizon form in the finite future of the collapsing matter themselves' time definition, their effects on the motion of the outside fixed position observers must also happen in the form of horizon and singularity naturally. The outside fixed position observers cannot see such objects appearance only because they adopt a bad coordinate system or time definition.
 
However, this popular belief makes two mistakes rather hiddely, (i) taking the global viewpoint of the horizon's existence as the viewpoint of local observers and (ii) taking the physics law's general coordinate invariance as the physical reality's observer independence. The physics law here refers to relations like $G_{\mu\nu}=8\pi G_NT_{\mu\nu}$ which bridge observables following from at least two types of physical measurement.  The physical reality refers to any physic quantity or physical features which is measured through only one type of physical measurement. These are two totally different concepts, the invariance of the physics laws under the general coordinate transformation does not imply the physical reality's observer independence in any sense. In fact, the physical reality's observer independence is not a fact at all. Let us elaborate this point through three examples. 

\begin{table}
\begin{tabular}{ll}
\hline
\multicolumn{2}{c}{$\begin{array}{l}ds^2{=}{-}hdt^2{+}h^{-1}dr^2{+}\cdots$,$h=1{-}\frac{2GM}{r}
\\k^\mu=(\frac{\omega_0h_0^{\!1\!/\!2}}{h},\omega_0h_0^{\!1\!/\!2},0,0)\end{array}$}
\\
$u^\mu_\mathrm{fx}{=}(h^{{-}{1\!/\!2}}_\mathrm{fx},0,0,0)$&$u^\mu_\mathrm{fr}{=}(\frac{\sqrt{h_0}}{h},\sqrt{h_0{-}h},0,0)$
\\
$\omega_\mathrm{fx}{=}{-}u_\mathrm{fx}{\cdot}k=\frac{\omega_0h_0^{\!1\!/\!2}}{h_\mathrm{fx}^{\!1\!/\!2}}$&$\omega_\mathrm{fr}{=}{-}u_\mathrm{fr}{\cdot}k{=}\frac{\omega_0[h_0{-}(h_0^2{-}h_0h)^{\!1\!/\!2}]}{h}$
\vspace{1mm}\\
\hline
\multicolumn{2}{c}{$\begin{array}{l}ds^2{=}{-}hdv^2{+}2dvdr{+}\cdots$,$h=1{-}\frac{2GM}{r}
\\
\bar{k}^\mu=(\frac{2\omega_0h_0^{\!1\!/\!2}}{h},\omega_0h_0^{\!1\!/\!2},0,0)\end{array}$}
\\
$\bar{u}^\mu_\mathrm{fx}{=}(h^{{-}{1\!/\!2}}_\mathrm{fx},0,0,0)$ &$\bar{u}^\mu_\mathrm{fr}{=}(\frac{\sqrt{h_0}{+}\sqrt{h_0{-}h}}{h},\sqrt{h_0{-}h},0,0)$
\\
$\bar{\omega}_\mathrm{fx}{=}{-}\bar{u}_\mathrm{fx}{\cdot}\bar{k}=\frac{\omega_0h_0^{\!1\!/\!2}}{h_\mathrm{fx}^{\!1\!/\!2}}$
&$\begin{array}{l}\bar{\omega}_\mathrm{fr}{=}{-}\bar{u}_\mathrm{fr}{\cdot}\bar{k}{=}\frac{\omega_0[h_0{-}(h_0^2{-}h_0h)^{\!1\!/\!2}]}{h}\end{array}$
\vspace{1mm}\\
\hline
\end{tabular}
\caption{Two types of observer's measurement of a light signal emitted from the surface of a freely collapsing star whose outside geometry is exactly Schwarzschild type. The upper is in the standard Boyer-Linderquist coordinate, the lower is in the Eddington-Finckelstein coordinate. As a physical reality, the light signal's frequency is invariant under general coordinate transformation but observer dependent. $h_0$ is the value of $h$ at the light signal's starting point. $h_\mathrm{fx}$ is the value of $h$ at the point the fixed position observer sits on. $k^\mu$ is the wave number of the light signal, whose components are determined by $\dot{k}^0+\Gamma^0_{\mu\nu}k^\mu k^\nu=0$ and $k{\cdot}k=0$. $u^\mu_\mathrm{fx}$ is the four velocity of the fixed position observer far away, $u^\mu_\mathrm{fr}$ is the four velocity of the freely falling observer outside the star. The components of $u^\mu$ are determined by $\dot{u}^0+\Gamma^0_{\mu\nu}u^\mu u^\nu=0$ and $u{\cdot}u=-1$}
\label{tableGCIandOBDexampleA}
\end{table}
The first is a concrete measurement example. Consider two observer's measuring of a same light signal's frequency emitted from an atom fixed on a freely collapsing star whose outside geometry is written in two different coordinate system. The first observer is a fixed position one, the second is a freely falling one. TABLE \ref{tableGCIandOBDexampleA} lists these two observer's definition and measurement mathematics. From the Table, we easily see that, the frequencies measured by the two observers $\omega_\mathrm{fx}$ and $\omega_\mathrm{fr}$ are obviously unequal, although they are invariant before and after the coordinate is transformed from the Boyer-Lindrquist to the Edington-Finkelstein type, $\omega_\mathrm{fx}=\bar{\omega}_\mathrm{fx}$, $\omega_\mathrm{fr}=\bar{\omega}_\mathrm{fr}$. In this example, the frequency of the light signal, and even the redshift's dependence on the radial coordinate $r$ is what we called physical realities. These realities are obviously observer dependent, because their value follows from the inner product of the observer's four velocity and the tensor representing the observable itself.

\begin{table}
\begin{tabular}{ll}
\hline\hline
\multicolumn{2}{c}{$\begin{array}{l}
ds^2{=}{-}hdt^2{+}h^{-1}dr^2{+}\cdots,h{=}1{-}\frac{2GM}{r}$,$2GM{\equiv}r_s
\\T^{\mu\nu}{=}\left[\begin{array}{cc}\rho\dot{t}\dot{t}&\rho\dot{t}\dot{x}\\\rho\dot{t}\dot{x}&\rho\dot{x}^2\end{array}\right], T^{\theta\theta}{=}0, T^{\phi\phi}{=}0\\h\dot{t}{\equiv}\gamma,\dot{x}^2{=}\gamma^2{-}h,\rho{=}\frac{m\delta(r{-}x)}{4\pi x^2},m{\ll}M
\end{array}$}
\\
\hline
$u^\mu_\mathrm{fx}=(h_\mathrm{fx}^{-{1/2}},0,0,0)$ & $u^\mu_\mathrm{co}{=}(\dot{t},\dot{x},0,0)$
\\
$\begin{array}{l}p_\mathrm{fx}=T^{\mu\nu}\frac{u^\mathrm{fx}_\mu u^\mathrm{fx}_\nu{+}g^\mathrm{fx}_{\mu\nu}}{3}
\\{=}\frac{\rho}{3}(h_\mathrm{fx}^{-1}\dot{t}^2{-}h_\mathrm{fx}\dot{t}^2{+}h^{-1}_\mathrm{fx}\dot{x}^2)
\\=\frac{\rho}{3}\frac{(h_\mathrm{fx}^{-1}{-}h_\mathrm{fx})\gamma^2}{h(x)^2}{+}\frac{\gamma^2{-}h}{h_\mathrm{fx}}
\\\xrightarrow{x\rightarrow r_s}\infty
\end{array}$ & $\begin{array}{l}p_\mathrm{co}{=}T^{\mu\nu}\frac{u^\mathrm{co}_\mu u^\mathrm{co}_\nu{+}g^\mathrm{co}_{\mu\nu}}{3}\\{=}{\rho}u^\mu_\mathrm{co}u^\nu_\mathrm{co}\frac{u_\mu^\mathrm{co}u_\nu^\mathrm{co}+g^\mathrm{co}_{\mu\nu}}{3}
\\{\equiv}0\\~\end{array}$
\\
\hline\hline
\multicolumn{2}{c}{$\begin{array}{l}d\bar{s}^2{=}{-}d\tau^2{+}\frac{r_s^{2/3}d\varrho^2}{[\frac{3}{2}(\varrho-\tau)]^{2/3}}{+}\cdots,r{=}\big[\frac{3}{2}(\varrho{-}\tau)\big]^\frac{2}{3}(2GM)^\frac{1}{3}\\
\begin{array}{l}\bar{T}^{\mu\nu}{=}\Bigg[\!\begin{array}{cc}\rho(\dot{t}{+}\dot{x})^2\!&\!\frac{\rho(\dot{t}{+}\dot{x})}{h}\frac{r_s\dot{t}{+}r\dot{x}}{\sqrt{r_sr}}\\\mathrm{symm}\!\!&\!\!\!\frac{\rho}{h^2}\frac{r}{r_s}(\frac{r_s}{r}\dot{t}{+}\dot{r})^2\end{array}\!\Bigg],\bar{T}^{\theta\theta}{=}\bar{T}^{\phi\phi}{=}0\end{array}\end{array}$}
\\
\hline
$\begin{array}{l}\bar{u}^\mu_\mathrm{fx}=(h^{-1/2}_\mathrm{fx},\frac{(r_s/r_\mathrm{fx})^{1/2}}{h^{3/2}_\mathrm{fx}},0,0)
\\\bar{p}_\mathrm{fx}{=}\bar{T}^{\mu\nu}\frac{\bar{u}^\mathrm{fx}_\mu\bar{u}^\mathrm{fx}_\nu{+}\bar{g}^\mathrm{fx}_{\mu\nu}}{3}
\\=\frac{\rho}{3}\frac{(h_\mathrm{fx}^{-1}{-}h_\mathrm{fx})\gamma^2}{h(x)^2}{+}\frac{\gamma^2{-}h}{h_\mathrm{fx}}
\\\xrightarrow{x\rightarrow r_s}\infty\end{array}$ 
&$\begin{array}{l}{\bar{u}}^\mu_\mathrm{co}{=}(\dot{t}{+}\dot{x},\frac{r_s\dot{t}{+}r\dot{x}}{h\sqrt{r_sr}},0,0)
\\\bar{p}_\mathrm{co}{=}\bar{T}^{\mu\nu}\frac{\bar{u}^\mathrm{co}_\mu \bar{u}^\mathrm{co}_\nu{+}\bar{u}^\mathrm{co}_{\mu\nu}}{3}\\{\equiv}0\\~\end{array}$
\\
\hline\hline
\end{tabular}
\caption{Two types of observers' measurement of the pressure of a dust type mass $m$ test spherical shell freely falling towards the horizon of a pre-existing Schwarszchild black hole of mass $M$. The upper is in the Boyer-Lindquist coordinate, the downer is in the Lemaitre coordinate. As a physical reality, the test mass shell's pressure is invariant under the general coordinate transformation, but observer dependent. $T^{\mu\nu}$ is the energy momentum tensor of the test mass shell; $u^\mu_\mathrm{fx}$ and $u^\mu_\mathrm{fr}$ are the four velocity of the fixed position observer and the freely falling observer respectively. $\bar{T}^{\mu\nu}$, $\bar{u}^\mu_\mathrm{fx}$ and $\bar{u}^\mu_\mathrm{fr}$ are their counter parts in the Lemaitre coordinate system.}
\label{tableGCIandOBDexampleB}
\end{table}

The second is still a concrete measurement example. But this time let us consider two observers measurement of a same test shell's pressure when it is freely falling towards the horizon of a pre-existing Schwarzschild black hole characterised by the exact horizon and central singularity. The first observer is a fixed position observer which is sitting far away from the horizon of the black hole. The second is a freely falling observer co-moving with the test shell. We display these two observers' four velocity and measurements of the test mass shell's pressure in TABLE \ref{tableGCIandOBDexampleB}. In this example, the pressures of the test mass shell is our physical reality. From the table, we can easily see that this physical reality is also observer dependent  $p_\mathrm{fx}\neq{}p_\mathrm{co}$, although coordinate invariant,  $p_\mathrm{fx}=\bar{p}_\mathrm{fx}$ and $p_\mathrm{co}=\bar{p}_\mathrm{co}$. $p_\mathrm{fx}$ has the property of approaching to infinite as the shell falls infinitely close to the horizon. So to the far away fixed position observers, the asymptotic mathematical horizon's existence makes the black hole look like an incompressible but finitely sized objects. $p_\mathrm{co}$ has the property of being zero constantly, which means that to the co-moving observer, the horizon exhibit not any observational effect at al.

\begin{figure}[ht]
\includegraphics[totalheight=55mm]{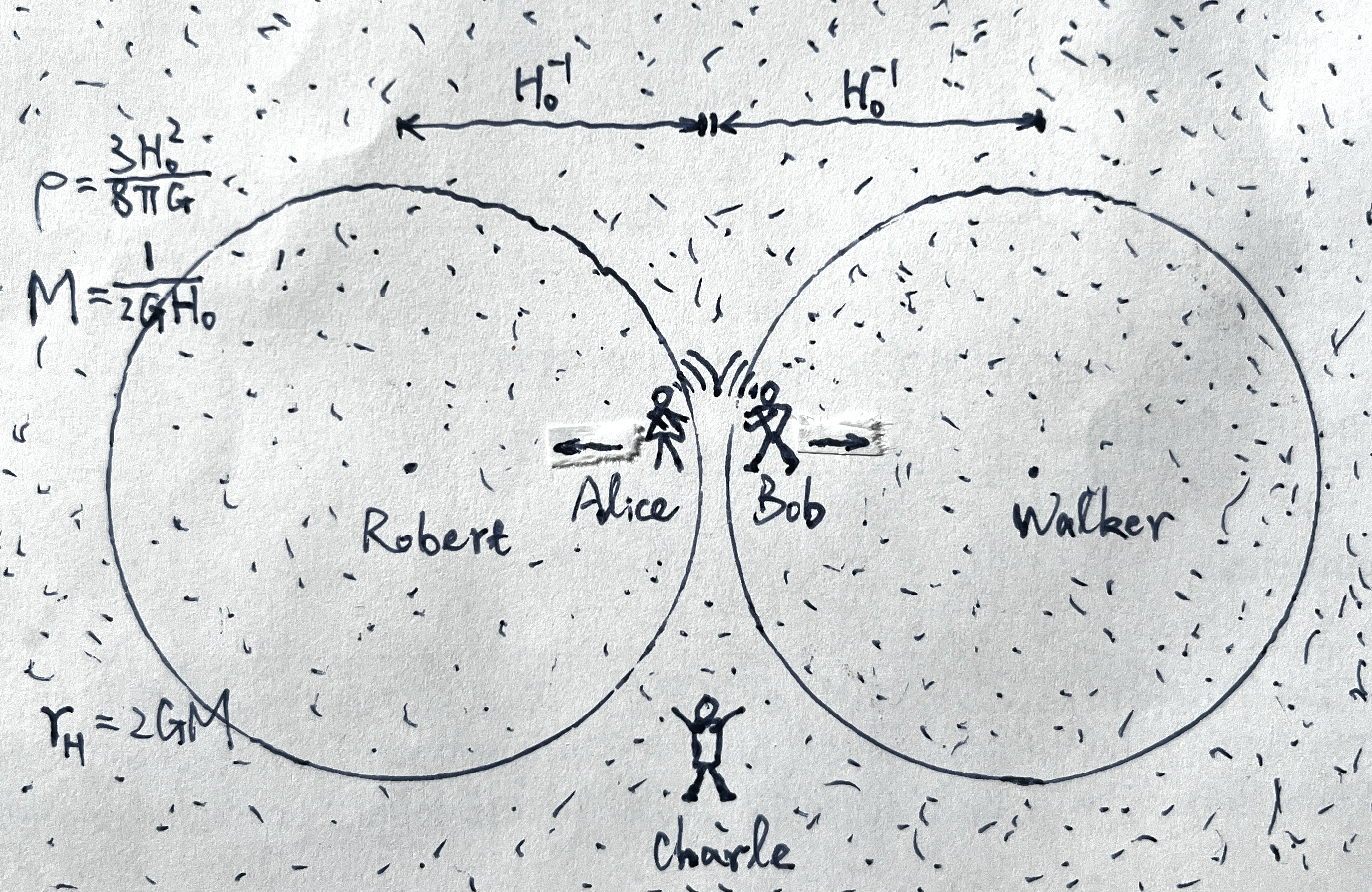}
\caption{The Face to Face Communication Difficulty (F2FCD). In an spatial flat and homogeneous Friedman-Robertson universe, Alice and Bob are a couple who date frequently. Behind each of them is their ex-boy/girl friend Robert and Walker respectively, two spherical giants of radius $H_0^{-1}$ and average mass density $\frac{3H_0^2}{8\pi G_N}$, $H_0$ denotes just the Hubble constant of the current universe. Although divorced for almost $1.37\times10^{10}$ years already, the gravitation between them and their own ex-friend still makes Alice and Bob recede from each other everyday. From Alice and Bob, this is nothing but the usual freely falling. Charle, their common friend or a global observer, also thinks this is nothing but the conventional big bang. The question is, why is it possible Alice and Bob get known to each other and begin to date? Because each of them lives inside the horizon caused by her or his ex-friend but outside that of the other's.}
\label{figF2Fcdp}
\end{figure}
The third is a conceptual example. Let us consider the thought experiment cartooned in FIG.\ref{figF2Fcdp} which contains a paradox we called F2FCD, see the figure and captions carefully. This paradox magnifies the error of the common belief that the horizon and singularity are observer independent physical realities to all black holes. Charle's confusion arises from two factors. (I) believing that in Alice's freely falling reference frame, Robert has a horizon and similarly in Bob and Walker's case. This is an error of taking the global viewpoint of Charle himself as the viewpoint of local observers such as Alice and Bob. (II) believing that the horizon-carrying feature of Robert to Alice is also the case to Bob and vice versa. This is an error of taking the physic laws' general coordinate invariance as the physical reality's observer independence.  In fact, as an outside observer, in any finite time of his own, Bob can not see or detect Robert's horizon at all. So he can get known with Alice without any difficulty.  This F2FCD paradox is firstly introduced in reference \cite{dfzeng2018b} to necessitate the horizon's fuzziness to resolve the information missing puzzle. But its value may be more right here to help understanding the physical reality's observer dependence.

As long as we know that the horizon-carrying is not a physical reality of black holes independent of observers, we can easily understand that the microscopic state picture of black holes is observer dependent. By our exposition in section \ref{secMetricOFPO},\ref{secMetricComObserver} and \ref{secQuantisation}, to the outside fixed position observer, the matter contents of physical black holes live inside a spatial region which is infinitely close to that wrapped by the mathematic surface $r_h=2GM_\mathrm{tot}$. But they live there just like some in-compressible and static fluid. Because inside that region, matters inside each sphere of smaller radius has the same feature. This is nothing but the concept of the old frozen star \cite{frozenStar1971,Buchdahl1959,chand1964a,chand1964b,bondi1964}, see references \cite{BrusteinFrozen2301,Brustein2304,Brustein2310} for recent investigations. Except that the in-compressibility now is understood as a result of the infinite gravitational time delaying instead of any physical exclusive forces. To the co-moving observer, these matters just freely falling towards the central point, over cross each other there and then oscillate periodically. But these inside horizon motion and oscillation happen in the future of infinitely far away future by the outside fixed position observers' physic time definition. They are physical realities only in the sense of parallel universe or statistic ensembles. Just as we showed above and in our earlier works \cite{dfzeng2017,dfzeng2018a,dfzeng2018b,dfzeng2020,dfzeng2021,dfzeng2022}, quantisation of these matters' fluctuation on the frozen star background or oscillations across the central point both lead to the right area law featured Bekenstein Hawking entropy, thus form equivalent and complementarity definitions for the microscopic state of black holes. 

\section{Banana shape deformation}
\label{secBanana}  

\begin{figure}[ht]
\includegraphics[totalheight=42mm]{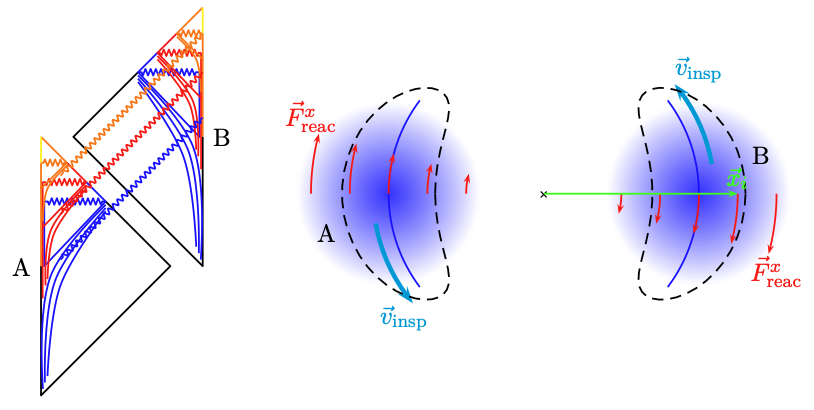}
\caption{When two black holes caused by gravitational collapse merge, both participants are outside observers of the other. So none of them can see the other's horizon as physical realities, instead each of them will receive light signal and thus gravitation from the matter contents of the other. The inhomogeneous back reaction of gravitational wave radiation would cause non-uniform orbit decay thus leading to banana shape-deformation in their matter contents' spatial distribution. The resulting black holes' quadrupole can be written as that of the two arc shape strings. }
\label{figBinaryBanana}
\end{figure}

It is very important to emphasise that the observers referred to in this work are not limited to human beings but are proxies of any probe or detectors, especially those living outside the matter occupation region of the being probed objects. They can be the imaging photons skimming over a single black hole or the other partner involved in a BHB merger event. These test objects' motion is determined by the gravitational field produced by the matter contents of the being tested black holes which never fully collapse and cause singularity in any finite future by the time definition of the outside fixed position observers. These being probed objects carry approximate horizon which is formed so long that during the short duration of the imaging process or binary merger process the imaging photon's perturbation or the inspiral partner's tidal force has no chances to cause disintegration. Because longer formation time means smaller matter occupation region radius thus more remarkable time delaying from the matter occupation region's boundary to the test particle's position, so that more time consuming for their matter contents to be affected by the imaging photon or merger partner's gravitation, and vice versa. In the BHB meger events, this provides us chances to treat dynamics of the binary system with the adiabatic approximation. By this approximation, we look the motion of the binary system as a combination of each participant's mass central's motion and their different parts' relative motion. The latter arises from the inhomogeneous radiation back reaction forces and will cause banana shape deformation of FIG.\ref{figBinaryBanana}.

The key reason for the banana shape deformation is the inhomogeneity of the gravitational wave radiation's back reaction. Referring to FIG.\ref{figBinaryBanana}, let us consider the newtonian motion of a representative point such as $\vec{x}_i$ on the participant B and its counter partner on participant A as an illustration for this phenomena. By the standard quadrupole formula for the back reaction force arising from the system's inspiral motion and gravitational wave radiation,
\bea{}
&&\hspace{-5mm}
F_i^\mathrm{reac}=\frac{G}{c^5}m_i\big\{-\frac{2}{5}x^i\frac{d^5Q_i}{dt^5}+\mathcal{O}[\frac{1}{c^2}]\big\}
\\
&&\hspace{-5mm}\frac{d^5Q_i}{dt^5}=\omega^5m_ix_i^2
\eea
we can write down the dynamic equation controlling the inspiral orbit's decay of these two representative point and get solutions
\bea{}
&&\hspace{-1mm}\frac{d}{dt}\big(-\frac{Gm_i^2}{2x_i}\big)=F_i^\mathrm{reac}\omega x_i
\Rightarrow\\
&&\hspace{-5mm}x_i=x_i^0(1-t/t_0)^\frac{1}{4},t_0=\frac{(x^0_i)^4c^5}{20G^3m_i^3}
\label{xEvolution}
\eea
The synchronicity of different composite points' inspiral motion $\frac{Gm_i^2}{4x_i^2}=m_i\omega^2x_i$ implies that $\frac{Gm_i}{4x_i^3}=\omega^2$ is an approximate constant for different point $i$. So as the inspiral motion progresses on those points with larger initial $x_i$ will reduce their orbit radius more remarkably
\beq{}
{\Delta}x=x_i^0-x_i^0(1-t/t_0)^{1/4},t_0=\frac{c^5}{1280\omega^6(x_i^0)^5}
\label{dxEvolution}
\eeq 
In the case of all points inspiral at the same angular frequency, this will cause the compression of the system along the $\vec{x}_i$ direction, the net effect is a banana shape deformation. The true case in general relativity would be much more complicated than the newtonian illustrations here and can only be simulated numerically. But the qualitative physical picture will be the same.

\begin{figure}[ht]
\includegraphics[totalheight=42mm]{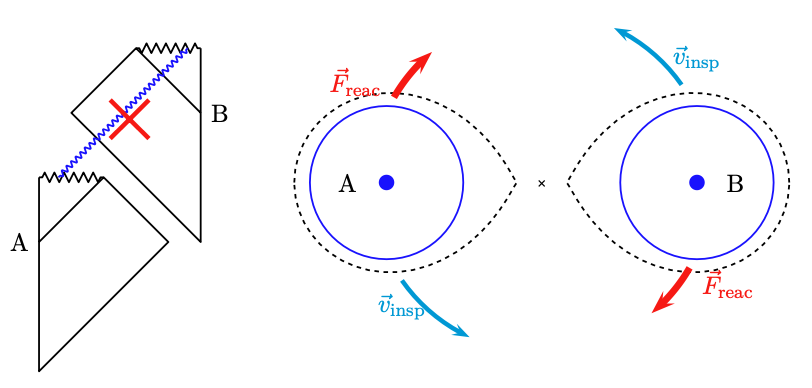}
\caption{When two black holes defined as exact solutions to the vacuum Einstein equation merge, both participants have absolute and apriori horizon. None of them can see the other's inside horizon structure. The back reaction force of the gravitational wave radiation can only affect the orbit of each other as a whole. None of them will experience inner-mass distribution's spatial change. Only their horizons will experience droplet shape deformation and merge into a single one finally. Considering the inhomogeneity of the time delaying effect, the true horizon shape deformation \cite{tidalDeformationNumeric} would be two comma like way of \raisebox{-3mm}{\huge`'}. }
\label{figBinaryDroplet}
\end{figure}

Obviously, when the two participating black holes experience banana shape deformation as in FIG.\ref{figBinaryBanana}, the radiation activity of the binary system will decrease inevitably, because in this case the rotational symmetry of the system enhances and the radiation relevant quadrupole of the system decreases correspondingly. The extremal case is when the two ends of the banana shape deformed black holes contact each other. In such cases the gravitational wave radiation will stop completely as the result of rotating symmetry's retrieving. For physic black holes originating from gravitational collapse, non-singular inner mass profile will make this banana shape deformation inevitable and will become very important as the two participants inspiral to the merger and ring down stage. From equation \eqref{dxEvolution}, we can see that this effect is of $\mathcal{O}[G^0]$, so it is non-perturbative and not accounted for in the higher multipole moments appearing as the response of the participating black holes to the tidal force caused by their partners and written into the tidal love numbers perturbatively. The tidal love numbers are well known to be zero for non-spinning black holes defined as exact solutions to the vacuum Einstein equation \cite{tidalLoveNumber2009Poisson,tidalLoveNumber2009Dmaour,tidalLoveNumber2015,tidalLoveNumber2021}. See FIG.\ref{figBinaryDroplet} for intuitive pictures.

\begin{figure}[h]
\begin{center}
\includegraphics[totalheight=30mm]{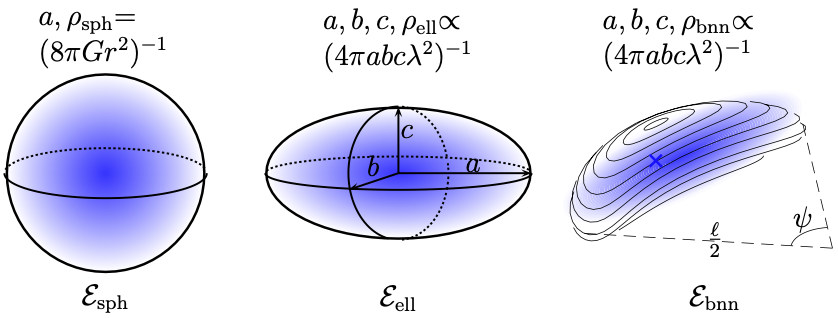}
\includegraphics[totalheight=25mm]{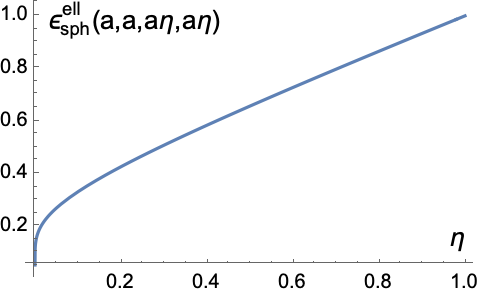}
\includegraphics[totalheight=25mm]{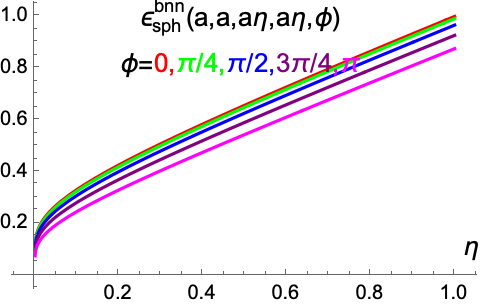}\end{center}
\caption{The ratio of self gravitational potential energy of ellipsoid and banana shape deformed star over that of the Schwarzshild black hole with equal surface potential $-GM/a=-\frac{1}{2}$ and $x$-directional size. The $y$,$z$-directional size are set to $b=c=a\eta$. The right is for banana shape deformation; $\psi$ is the bending angle opened by the arc of the banana relative to the radius of inspiral motion.}
\label{figSelfEnergy}
\end{figure}

Since both of the merger participants are outside observers of their inspiral partner, neither of them would see the other's horizon. Both of them would see the other as some frozen star occupying invariant spatial volume. So their compression along $\vec{x}_i$ direction implies stretching along the $\vec{v}_\mathrm{insp}$ direction inevitably. In a sense, the spatial volume contained in a spherical symmetric Schwarzshild horizon is the minimal volume we can compress matter contents into. This can be seen from the following fact that a spherically symmetrical collapsings star with linear radial mass profile $M(r)\approx\frac{r}{2G}$ implement the lowest possible self gravitating potential energy without causing the physical horizon's formation
\bea{}
&&\hspace{-9mm}\mathcal{E}_{sph}{=}-\!\!\int_0^a\!\!\frac{GM(r)\rho_\mathrm{sph}(r)4\pi r^2dr}{r}[M(r){=}\frac{r}{2G}]
\\
&&\hspace{-9mm}=-\frac{GM^2_\mathrm{tot}}{a},a=2GM_\mathrm{tot}, 
\eea
Note that $M(r)\approx\frac{r}{2G}$ implies $\rho_\mathrm{sph}(r)\approx\frac{1}{8\pi Gr^2}$ necessarily. Compressing this star and tuning its mass value so that the surface of the resulting ellipsoid has equal newtonian potential \cite{strattonEMTbook} as the spherical symmetric star $-\frac{GM}{a}=-\frac{1}{2}$, the self gravitation energy of the resultant object will become
\begin{align}
&\mathcal{E}_{ell}=-\!\int_0^1\!\frac{GM(\lambda)F(k,\varphi)\rho_\mathrm{ell}(\lambda)d\mathbf{v}}{\lambda\sqrt{a^2-c^2}}
=-\frac{GM^2F(k,\varphi)}{\sqrt{a^2-c^2}}
\label{selfgravitatingenergyEllipsoid}\\
&k\equiv\frac{\sqrt{a^2{-}b^2}}{\sqrt{a^2{-}c^2}},\varphi\equiv\arcsin\sqrt{1{-}\frac{c^2}{a^2}}
\end{align}
where $F(k,\varphi)$ is the elliptic integral of the first kind and $d\mathbf{v}=abc\lambda^2d\lambda\sin\theta d\theta d\phi$. In this calculation, the star is considered as many ellipsoid shape shells of equal axis length ratio \{$a\lambda,b\lambda,c\lambda$\} and radial mass function $M(\lambda)=M\cdot\lambda$ or $\rho_\mathrm{ell}(\lambda)=\frac{M}{4\pi abc\lambda^2}$. By our convention, the star is compressed along $y$ and $z$ direction but kept invariant along the $x$-direction. Letting the surface potential of the resulting ellipsoid take the same value as that of the Schwarzschild black hole $\frac{GMF(k,\varphi)}{\sqrt{a^2-c^2}}=\frac{GM}{a}=\frac{1}{2}$ with the same $x$-directional size, we will find that $M_\mathrm{ell}<M_\mathrm{sph}$ and $\mathcal{E}_\mathrm{ell}/\mathcal{E}_\mathrm{sph}<1$.  See FIG.\ref{figSelfEnergy} for concrete numerics. Conversely, it is very natural to understand that if we keep the total mass invariant when compressing the star along one direction, then the star must be stretched along the other so that the horizon formation condition $\frac{GMF(k,\varphi)}{\sqrt{a^2-c^2}}=\frac{1}{2}$ will not be realised as any outside test body cannot detect it in the finite future by their physical time definition.

Considering the banana shape deformation, we will get similar conclusion. In the case the ellipsoid star are parameterised as, 
\bea{}
&\hspace{-7mm}x{=}a\lambda\sin\theta\cos\phi,y{=}b\lambda\sin\theta\sin\phi,z{=}c\lambda\cos\theta
\\
&\hspace{-7mm}\lambda\in(0,1),\theta\in(0,\pi),\phi\in(0,2\pi)
\eea
the banana shape deformation can be written as
\bea{}
&\hspace{-5mm}x=(\frac{\ell}{2}{+}a\lambda\sin\theta\cos\phi)\cdot\cos\big(\frac{b\lambda\sin\theta\sin\phi}{\frac{\ell}{2}+a\lambda\sin\theta\cos\phi}\big)
\\
&\hspace{-5mm}y=(\frac{\ell}{2}{+}a\lambda\sin\theta\cos\phi)\cdot\sin\big(\frac{b\lambda\sin\theta\sin\phi}{\frac{\ell}{2}+a\lambda\sin\theta\cos\phi}\big)
\\
&\hspace{-5mm}z=c\lambda\cos\theta
\eea
Similar with the ellipsoid shape deformation case, still looking the banana as many concentric banana-peel whose own volume has the form $4\pi a_\mathrm{bnn}b_\mathrm{bnn}c_\mathrm{bnn}\lambda^2d\lambda$ and wrap mass $M\cdot\lambda$ inside, then the self-gravitational potential energy of the system can be calculated approximately as
\begin{align}
&\hspace{-5mm}\mathcal{E}_\mathrm{bnn}=-\int_0^1\frac{Gm(\lambda)\rho_\mathrm{bnn}(\lambda)d\mathbf{v}_\mathrm{bnn}}{\lambda\sqrt{a^2-c^2}}F_\mathrm{bnn}(k,\varphi)
\\
&\hspace{-5mm}\approx-\frac{GM^2}{\sqrt{a^2_\mathrm{bnn}-c^2_\mathrm{bnn}}}F(k_\mathrm{bnn},\varphi_\mathrm{bnn})
\label{bnnPotentialApproximate}
\\
&\hspace{-5mm}k_\mathrm{bnn}\equiv\frac{\sqrt{a^2_\mathrm{bnn}{-}b^2_\mathrm{bnn}}}{\sqrt{a^2_\mathrm{bnn}{-}c^2_\mathrm{bnn}}},\varphi\equiv\arcsin\sqrt{1{-}\frac{c^2_\mathrm{bnn}}{a^2_\mathrm{bnn}}}
\\
&\hspace{-5mm}a_\mathrm{bnn}=\frac{a\sin{\psi/2}}{\psi/2},b_\mathrm{bnn}=b,c_\mathrm{bnn}=c
\end{align}
where $\psi$ is the bending angle of the banana, referring to FIG \ref{figSelfEnergy} for physical pictures; subscripts $\scriptscriptstyle\mathrm{bnn}$ means banana shape deformation. In the case of $\psi\ll\pi$, equation \eqref{bnnPotentialApproximate} provides rather good approximation to the underlying exact result. For large $\psi$, equation \eqref{bnnPotentialApproximate} is not a good approximate, but still catches the qualitatively feature of banana shape deformation costing extra energy relative to the ellipsoid deformation.

\begin{figure}[h]
\begin{center}
\includegraphics[totalheight=40mm]{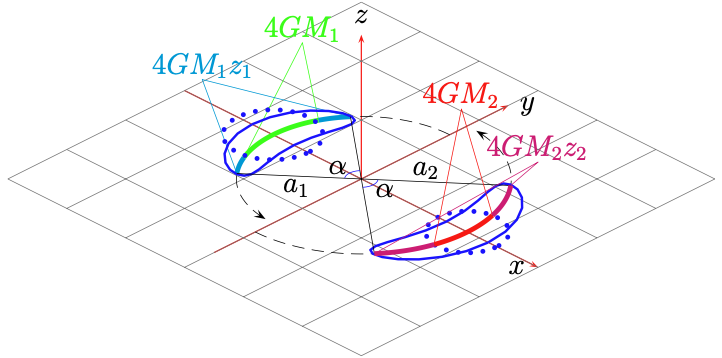}\end{center}
\caption{The radiation active quadrupole of a banana shape deformed BHB system is suppressed by a factor of $\zeta=\frac{\sin\alpha}{\alpha}$. $4GM_i$ and $4GM_iz_i$ are the deformation before and after diameter/length of the black holes respectively.}
\label{figBnnFactor}
\end{figure}

Referring to FIG.\ref{figBnnFactor}, when we consider a banana shape deformed BHB system's rotation and gravitational wave radiation, the time dependent quadrupole will become
\beq{}
Q_{xx}=\sum_{i=}^{1,2}\!\!\int_{-\frac{\alpha}{2}+0/\pi+\omega t}^{\frac{\alpha}{2}+0/\pi+\omega t}\!\!\frac{M_i}{\alpha a_i}(a_i\cos\theta)^2a_id\theta
\label{QxxDerivation}
\eeq
\beq{}
=\big(M_1a_1^2{+}M_2a_2^2\big)\frac{\sin\!\alpha\cos\!2{\omega}t{+}\alpha}{2\alpha}\sim\mu a^2\frac{\sin\!\alpha\cos2{\omega}t}{2\alpha}
\nonumber
\eeq
\beq{}
Q_{yy}{=}\sum_{i=}^{1,2}\!\!\int\!\!\frac{M_i}{\alpha a_i}(a_i\sin\theta)^2a_id\theta{\sim}{-}\mu a^2\frac{\sin\!\alpha\cos2{\omega}t}{2\alpha}
\nonumber
\eeq
where $\mu\equiv\frac{M_1M_2}{M}$ and $a_{1,2}{=}\frac{M_{2,1}a}{M}$, $M{=}M_1{+}M_2$. To assure the shape deformed black holes open equal angles $\alpha_1\equiv\frac{4GM_1z_1}{a_1}=$ $\frac{4GM_2z_2}{a_2}\equiv \alpha_2=\alpha$ with respect to their inspiral central, the elongation factor $z_i$ is required to satisfy $\frac{z_1M_1}{M_2}=\frac{z_2M_2}{M_1}\equiv z$. As results, the radiation active quadrupole of the system can be written as
\beq{}
Q^\mathrm{radiation}_\mathrm{active}=\zeta\cdot\mu a^2,
\zeta=\frac{\sin\!\alpha}{\alpha}=\frac{\sin\!4GMz/a}{4GMz/a}
\label{zetaDefinition}
\eeq
Just as we pointed out previously, during the inspiral and merger process the banana shape deformed black hole would be elongated under the inhomogeneous radiation back reaction force, so $z$ is a growing function of the time coordinate $t$. In this work, we will not try to determine the function form of $z(t)$ dynamically but choose to parameterise it as various power functions of $a(t)$
\beq{}
z[a(t)]=0,1,a(t)^{-1},a(t)^{-2},\cdots,
\label{zaParamerize}
\eeq
As time passes by, $a(t)$ would decrease due to the relative motion orbit decay caused by gravitational wave radiation. So $a(t)^{-n}(0{<}n)$s are increasing functions of time. $z=0$ corresponds to the case the black hole has point like singular structure. $z=1$ corresponds to the case the black hole has extended inner structure but such structures experience only banana shape bending but no elongation as the system inspirals and emits gravitational wave radiation. Black holes with $z{\propto}a(t)^{-1}$ experience more banana shape elongation during the early inspiral stage than those with $z{\propto}a(t)^{-2}$ so are more soft than the latter. It is necessary to note that although the black holes with non-singular inner-structure experience banana shape deformation, the protection provided by their asymptotic horizon would prevent them from tidal disintegration during the most duration of their binary merger process, from the early inspiral to the very late ring-down stage. However, to show this inner structure truly exists and happens there, we need to develop new theoretical tools to calculate their observational signal analytically.

\section{An exact one body method}
\label{secXOBmethod}

The full process of a BHB system's evolution can be divided into three stages, i.e. inspiral, merger and ring-down. The post newtonian approximation and effective one body (EOB) method can be used to describe evolutions of the system during the first and early part of the second stage. Numeric relativity can be used to model the system's evolution during the whole process but is usually applied only for the second stage due to computational resource contstraint. The practical signal wave form used in observation are compositions of the three method's output, with the joining condition tuned carefully by hand.

Even one day, our computational resource gets rich enough to calculate the wave form of the merger process with numeric method exclusively, the popular belief that Schwarzschild black hole has singular inner-mass distribution is still not a conclusion extractable from such calculations. The reason is, Kerr black holes have extended inner structure embodied in their singular ring. If such a structure follows from the merger events of two Schwarzschild black holes, then the inner mass distribution of the parent Schwarzschild black hole must not be point-like singular. Instead it must be extended and will experience stretching and bending during the binary merger process. Signals of such structures would be hidden inside the boundary data \cite{NR2002,NR2004,NR2007,NR2018} of numeric relativity, what we see in the inner-motion irrelevance is only an illusion. To understand the existence of such an inner-structure picture observationally, what we should do is not to escalate the precision of the existing numeric relativity's simulation by finding more precise initial data basing on higher order post newtonian expansions or something else. What we should do is to develop some more self consistent and full process applicable analytical description or analytical wave form generating techniques which directly translate the inner-structure of black holes to the gravitational wave forms measurable from the real BHBs' merger events.

\begin{figure}[ht]
\includegraphics[totalheight=32mm]{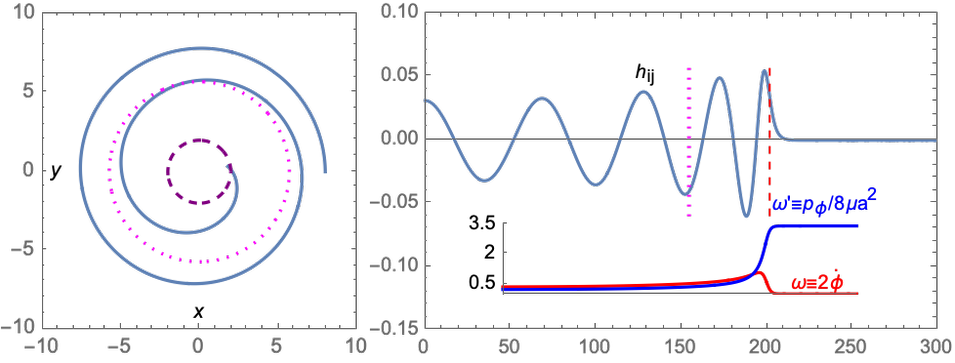}
\caption{The relative motion orbit of general relativity two particle system and the corresponding gravitational wave form following from the conventional EOB method. This method yields manifestly opposite variation trends for the system's inspiral frequency defined through $\omega{\equiv}\dot{\phi}$ and $\omega{\sim}p_\phi/\mu a^2$ as the system evolves to the very late stage.}
\label{figEOBinconsistence}
\end{figure}

The basic idea of conventional EOB method \cite{eob1998,eob2000,eob2009} is to map the general relativity two bodies' motion onto those of a total mass particle's shifting and a reduced mass particle's inspiral. Due to the non-linearity of general relativity, the gravitational field controlling the reduced mass particle's inspiral is not a Schwarzschild metric sourced by the total mass particle, but an effective and corrected but still static and Schwarzschild-like one. This effective metric must be constructed by requiring that the reduced mass particle's motion in it given by the world line action coincide with the two particle's relative motion in the original general relativity two-body problem under the post newtonian approximation. For the inspiral and early merger stage, this method can yield gravitational wave forms coincide with those following from numeric relativities very precisely. However, as the late merger and ring-down stage is coming, this method's inconsistence manifests. Most manifestly, the relative inspiral motion's frequency following from $\omega{\equiv}\dot{\phi}$ and $\omega{\sim}p_\phi/\mu{a^2}$ exhibits opposite trends of variation as the time passes by, see FIG.\ref{figEOBinconsistence} for illustrations. This means that this method cannot be used to tell us wheter the two Schwarzschild black holes involved in the binary merger process own inner-structure and experience shape deformation or not.

\begin{figure}[ht]
\includegraphics[totalheight=40mm]{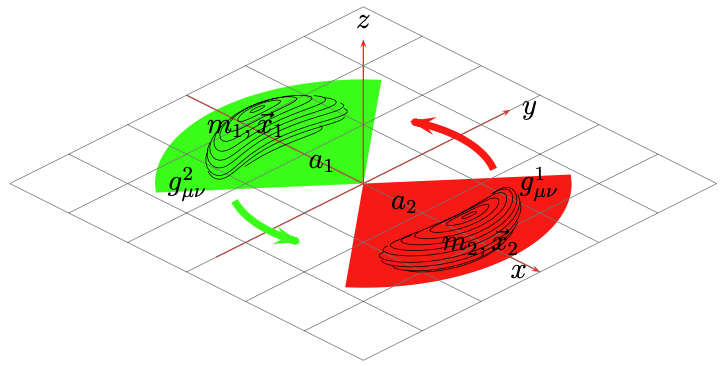}
\caption{The spacetime is dynamically partitioned into three regions, red, green and their inter-connection region. Each region has its own static geometry whose application region rotates synchronously with the merger participants. For our purposes, we only need to know the metric functions' form in the red and green region and in the equatorial plane.}
\label{figRWmetric}
\end{figure}

The conventional EOB method's inconsistency arises from its decomposition of the essentially time dependent radiation geometry into ``a static background gravitation + dissipative test particle''. This doing is not self-consistent because a test particle's motion in a fixed background spacetime is well known to be conservative so no radiation caused dissipation is allowed at all. To avoid this inconsistency, we come back to the original general relativity two-body action
\beq{}
S{=}{-}\sum_{i=}^{\scriptscriptstyle 1,2}\!\!\int\!\!{d\tau}\sqrt{-||g_{\mu\nu}(X_i)\dot{X}^\mu_{i}\dot{X}^\nu_{i}||}{-}\!\!\int\!{F}^\mathrm{diss}{d\tau}
\label{worldlineAction}
\eeq
and write the time dependent radiation geometry as the combination of three independent static patches\footnote{To determine the binary system's inspiral motion, what we need is only the space-time metric on the motion plane instead of the whole space. So exactly speaking, the metric expressions \eqref{rwmetric1}-\eqref{rwmetric2} are valid only on the motion plane.} with time-dependent application region, see FIG.\ref{figRWmetric} for intuitive pictures. On the motion plane, the two patches' accommodating the two merger participants and rotating synchronously with them can be written as, 
\beq{}
g^1_{\mu\nu}{=}\mathrm{diag}\{1{-}\frac{2GE_2}{a_1},(1{-}\frac{2GE_2}{a_1})^{-1},a_1^2,a_1^2\sin^2\!\theta_1\}
\label{rwmetric1}
\eeq
\beq{}
g^2_{\mu\nu}{=}\mathrm{diag}\{1{-}\frac{2GE_1}{a_2},(1{-}\frac{2GE_1}{a_2})^{-1},a_a^2,a_2^2\sin^2\!\theta_2\}
\label{rwmetric2}
\eeq
\beq{}
X_1^\mu{=}\{\tau,a_1(\tau),\frac{\pi}{2},\phi_1(\tau)\},X_2^\mu{=}\{\tau,a_2(\tau),\frac{\pi}{2},\phi_2(\tau)\}
\eeq
By this application-region wise metric, the action of the general relativistic two-body system becomes
\bea{}
S{=}{-}M_1\!\!\int\!\!d\tau\!\sqrt{1{-}\frac{2GE_2}{a_1}{-}(1{-}\frac{2GE_2}{a_1})^{-1}\dot{a}_1^2-a_1^2\dot{\phi}_1^2}
\rule{5pt}{0pt}\label{worldlineActionSolidified}
\\
{-}M_2\!\!\int\!\!d\tau\!\sqrt{1{-}\frac{2GE_1}{a_2}{-}(1{-}\frac{2GE_1}{a_2})^{-1}\dot{a}_2^2-a_2^2\dot{\phi}_1^2}
\rule{5pt}{0pt}
\nonumber
\\
+S_\mathrm{diss}
\rule{5pt}{0pt}
\nonumber
\eea
Since the orbit decay caused by gravitational wave radiation is always slow relative to the rotation of the system, we can always neglect the variation of the two participants' radial coordinate in a single circular orbit motion period thus setting $\dot{a}_1=\dot{a}_2=0$. So for the circular orbit motion, the variation principle for $a_1$ and $a_2$ will yield that
\beq{}
\omega^2_1\equiv\dot{\phi}^2_1=\frac{GE_2}{a_1^3},\omega^2_2\equiv\dot{\phi}^2_2=\frac{GE_1}{a_2^3}
\label{kepplerThirdLawA}
\eeq
For the non-circular orbit motion, as long as we understand $a_1$ and $a_2$ as the length of the semi-major axis of the two participants' approximately periodic elliptic orbit, equation \eqref{kepplerThirdLawA} will also be valid. At the same time, to assure the unshifting of the two merger participants' central of mass position, we impose
\beq{}
a_1=\frac{M_2}{M}a,a_2=\frac{M_1}{M}a, M=M_1{+}M_2
\label{a1a2relation}
\eeq
As results, the synchronicity of the two participants' inspiral motion implies that
\beq{}
\omega_1=\omega_2\Rightarrow E_2=\frac{M_1^3}{M^2}, E_1=\frac{M_2^3}{M^2}
\label{E1E2expression}
\eeq
It can be easily verified that, by this choice of $E_1$ and $E_2$, equations \eqref{kepplerThirdLawA} will lead to a general relativistic version of Keppler's third law for the reduced mass particle's motion in the total mass particle's gravitational field. 

Equations \eqref{a1a2relation} and \eqref{E1E2expression} transform the two-body problem \eqref{worldlineActionSolidified} into a single body one
\bea{}
L{=}{-}M_1\big[1{-}\frac{2GM}{a}{-}(1{-}\frac{2GM}{a})^{-1}\frac{M_2^2}{M^2}\dot{a}^2{-}\frac{M_2^2}{M^2}{a}^2\omega^2\big]^\frac{1}{2}
\nonumber\\
{-}\!M_2\big[1{-}\frac{2GM}{a}{-}(1{-}\frac{2GM}{a})^{-1}\frac{M_1^2}{M^2}\dot{a}^2{-}\frac{M_1^2}{M^2}{a}^2\omega^2\big]^\frac{1}{2}
\nonumber
\\
{+}L_\mathrm{diss}\rule{10mm}{0pt}
\label{singlebodyAction}
\eea
Although destined to be descriptions for the relative motion of the two bodies, this lagrangian differs from that of the reduced mass particle in the background Schwarzschild field of the total mass particle, even in the equal mass and exactly circular orbit case, because $\frac{M_2^2}{M^2}$ or $\frac{M_1^2}{M^2}$ is always less than $1$. 
The logic of the one body-lisation method \eqref{singlebodyAction} is to capture the non-linearity of general relativity gravitation non-perturbatively through the non-perturbative central fixing condition \eqref{a1a2relation} and the relative motions' synchronicity condition \eqref{E1E2expression}. In contrasts, the logic of conventional EOB method of \cite{eob1998,eob2000,eob2009} is to capture the non-linearity of general relativity gravitation through the perturbative match of the relative motion hamiltonian with that of a test body's motion in the background of a static and corrected Schwarzschild-like field. Since our one body-lisation conditions \eqref{a1a2relation}-\eqref{E1E2expression} are exact and call no post-newtonian approximation as input, we name it as an eXact One Body (XOB) method.

For those who can't help from evaluating how far is our one body-lisation idea to the conventional EOB method, the post newton expansion of the hamiltonian following from the conservative part of \eqref{singlebodyAction} is as follows
\beq{}
H\equiv\frac{1}{\mu}\big[\dot{a}\frac{\delta L}{\delta\dot{a}}+\omega\frac{\delta L}{\delta\omega}-L]
,\mu{\equiv}\frac{M_1M_2}{M_1{+}M_2},M{\equiv}M_1{+}M_2
\label{singlebodyHamiltonian}
\eeq
\beq{}
H_0=\frac{1}{\nu}+\frac{p^2}{2}-\frac{1}{q},p\equiv\dot{a},q\equiv\frac{a}{GM},\nu{\equiv}\frac{\mu}{M}
\label{postNewtonianHamiltonian0}
\eeq
\beq{}
H_2=(1-3\nu)\big[\frac{3p^4}{8}+\frac{3p^2}{2q}-\frac{1}{2q^2}\big]
\label{postNewtonianHamiltonian2}
\eeq
\beq{}
H_4=(1-5\nu+5\nu^2)\big[\frac{3p^6}{16}+\frac{21p^4}{8q}+\frac{15p^2}{4q^2}-\frac{1}{2q^3}\big]
\label{postNewtonianHamiltonian6}
\eeq
\beq{}
H_6{=}(1{-}7\nu{+}14\nu^2{-}7\nu^3)\big[\frac{35p^8}{128}{+}\frac{55p^6}{16q}{+}\frac{189p^4}{16q^2}{+}\frac{35p^2}{4q^3}{-}\frac{5}{8q^4}\big]
\label{postNewtonianHamiltonian8}
\eeq
Except $H_0$, all the higher order terms do not coincide with the input of the conventional EOB method \cite{eob1998,eob2000,eob2009}. However, just as we pointed out above and illustrated in FIG.\ref{figEOBinconsistence}, including the dissipation term in the hamiltonian of the conventional EOB method is not self-consistent. The inspiral frequency $\omega\equiv\dot{\phi}$ and $\omega{\sim}p_\phi/\mu{}a^2$ following from such a method exhibit manifestly opposite variation trends at the late time of the merger process. In contrast, adding dissipation term in our hamiltonian \eqref{singlebodyHamiltonian} is self consistent, because the frequency $\omega$ it involves plays the role of the merger participants' inspiral speed and the resultant spacetime's rotational speed simultaneously. So the whole system allows radiation caused dissipation consistently. 

\begin{figure}[ht]
\includegraphics[totalheight=32mm]{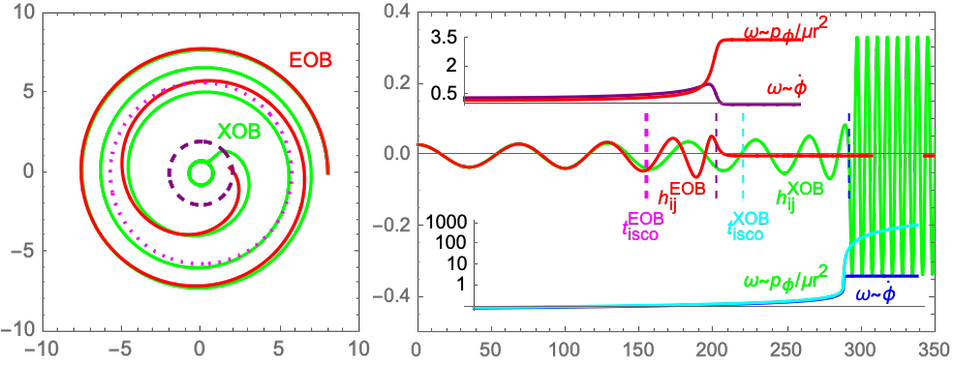}
\caption{The relative motion orbit and gravitational wave form of an equal mass BHB system following from the standard EOB method and our XOB method. Both members of the binary system are not inner-structured and experience not any shape deformation during the merger process. The dotted magenta line marks the ISCO radius defined by the EOB method. The dashed red line is the horizon radius of the total mass black hole $r_h=2GM_{tot}$.}
\label{figWFcomparing}
\end{figure}

For the above reason, we will not take the higher order post newton expansion of \eqref{postNewtonianHamiltonian2}-\eqref{postNewtonianHamiltonian8} differs from the input of reference \cite{eob1998,eob2000,eob2009} as a negative diagnosis for our idea's rationality. Instead we will take the direct gravitational wave form output of our method as the justification for its validity. Taking the most simple quadrupole radiation power as source of dissipation, we will get evolution equations for $a(t)$
\beq{}
\frac{dH}{da}\frac{da}{dt}=F_\mathrm{diss}=-\frac{32}{5}G\mu^2a^4\omega^6
\eeq 
From this equation we can solve the function $a(t)$ and calculate the gravitational wave through
\beq{}
h_{ij}\propto G\mu a^2\omega^3\cos2{\int\!\omega}dt
\label{gwfQuadrupole}
\eeq
We compare in FIG.\ref{figWFcomparing} the relative motion orbit and gravitational wave forms following from the EOB method and our XOB method. From the figure, we can easily see that the two orbits and wave forms highly agree with each other till the inner most circular orbit (ISCO) epoch of EOB. However, XOB exhibits much rational or self-consistent behaviour after that epoch. For example by XOB, the late time inspiral speed defined through $\omega=\dot{\phi}$ and $\omega\sim p_\phi/\mu a^2$($p_\phi{\equiv}\frac{\delta L}{\delta\omega}$) are both growing functions of time. While the variation trends of those predicted by EOB are contrary with each other. 

Two very impressive features of the gravitational wave forms following from XOB displayed in FIG.\ref{figWFcomparing} are (i) the decay of the gravitational wave forms is invisible from the time interval displayed and (ii) the asymptotic frequency $\omega{\sim}p_\phi/{\mu}a^2$ of the inspiral motion is logarithmically divergent. Both features imply that the binary system has infinite energy to be carried away by the gravitational wave radiation. This is provided by the infinite decreasable space of the gravitational potential energy between the two point particle type participating black holes. In the language of newtonian mechanics, the is just the reflection of $\varepsilon{=}-\frac{GM_1M_2}{a}\xrightarrow{a\rightarrow0}{-}\infty$. While in the language of general relativity, $\varepsilon{\propto}\frac{-GM_1M_2^2/(Ma)}{\sqrt{1-3GM_2^2/(Ma)}}{+}(M_1{\leftrightarrow}M_2)\xrightarrow{a\rightarrow a_f}{-}\infty$, where $a_f=\mathrm{max}\{\frac{3GM_1^2}{M},\frac{3GM_2^2}{M}\}$ is the asymptotical separation the two black holes can get close to, see equation \eqref{HbinaryMassB} of next section for references. In the case of $M_1=M_2$, this asymptotic value $a_f=\frac{3GM}{4}$ is less than the mathematic horizon radius $r_h=2GM$. 

From FIG.\ref{figWFcomparing}, careful readers may see that XOB method has the power of traceing relative motions of the binary system and apply the quadrupole formula \eqref{gwfQuadrupole} for the gravitational wave forms even when $a_f{<}2GM_\mathrm{tot}$, i.e. when the two participating black holes enter into the horizon determined by their total mass. This is rational because to each of them, their partner's horizon is only $2GM_i$ which is less than $a_f$. So all of them are outside observers of the other during the whole inspiral, merger and ring-down process. The horizon defined by $r_h=2GM_\mathrm{tot}$ is never implemented at all. So, tracing relative motions of the binary system inside its mathematical horizon does not violate any physical law such as causality at all. In a sense, this is a conceptual revolution. If one cannot understand this properly, he/she will have difficulties to understand why the outside fixed position observers can see the gravitational wave forms reflecting the relative motion of the two black holes inside the horizon defined by their total mass  $r_h\equiv2GM_\mathrm{tot}$. This will be even more clear when we consider the non-singular inner structure and banana shape deformation of the participating black holes. In such cases, XOB's ability to trace the system's evolution inside the mathematic horizon will bring us the full gravitational wave forms with well behaved late time quasi-normal mode feature and decodable information about the inner structure of the participating black holes.

By our XOB method,  during the BHB merger process the event of exiting from innermost stable circular orbit (ISCO) does not happen on the  epoch defined by the EOB method $a=5.718GM_\mathrm{tot}$. Because, by XOB method even after that epoch, the relative motion orbit of the system is stable and the system's evolution after the horizon entrance $r<2GM_\mathrm{tot}$ is still traceable analytically until the system enters the ring down phase. However, for comparing and referring convenience, we will preserve this nomenclature and take $5.718GM_\mathrm{tot}$ as $a_{\scriptscriptstyle\!ISCO}$, and mark it out in most of the relative motion orbit and gravitational wave form picture below.

\section{The full gravitational wave forms}
\label{fullWFinnerstructure}

Consider the non-singular inner-structure of black holes and their radiation activity suppression caused by banana shape deformation, we have to write the relative motion orbit decay equation as
\beq{}
\frac{da}{dt}=-\frac{32}{5}G\mu^2a^4\omega^6\zeta^2\cdot\big(\frac{dH}{da}\big)^{-1},\zeta=\frac{\sin4GMz/a}{4GMz/a}
\label{dadtGeneralExpression}
\eeq
where $\zeta$ is the radiation activity factor of the binary system and $z$ grows with time slowly, see equation \eqref{zetaDefinition} and comments there for their definition and concrete expressions. Focusing on the circular orbit or understanding $a$ only as the semi-major axis length of the quasi-periodic elliptic orbit, 
\beq{}
H[a,\nu,\omega]{=}\frac{M_2(1-\frac{2GM_1^2}{Ma})}{\sqrt{1{-}\frac{2GM_1^2}{Ma}{-}\frac{M_1^2\omega^2a^2}{M^2}}}
{+}\frac{M_1(1-\frac{2GM_2^2}{Ma})}{\sqrt{1{-}\frac{2GM_2^2}{Ma}{-}\frac{M_2^2\omega^2a^2}{M^2}}}
\label{HbinaryMassA}
\eeq
During each single quasi periodic evolution period the rotational speed of the system $\omega^2=\frac{GM}{a^3}$ will be approximately constant and the hamiltonian $H$ will become the function of $a$ and $\nu$ exclusively. 
\bea{}
&&\hspace{-5mm}H[a,\nu]^{\omega^2\approx}_{GM{\!/\!}a^3}{=}\frac{M_2(1{-}\frac{2GM_1^2}{Ma})}{\sqrt{1{-}\frac{3GM_1^2}{Ma}}}
{+}\frac{M_1(1{-}\frac{2GM_2^2}{Ma})}{\sqrt{1{-}\frac{3GM_2^2}{Ma}}}
\label{HbinaryMassB}
\eea 

\begin{figure}[ht]
\includegraphics[totalheight=50mm]{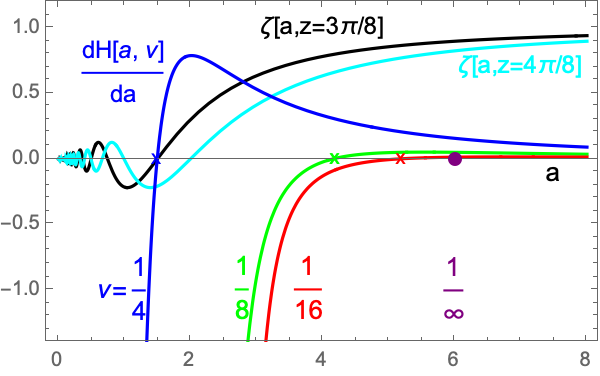}
\caption{The zero point of $dH[a,\nu]/da$ happens between $\frac{3}{2}{<}a{<}6$. The lower bound happens on $\nu=1/4$, the upper bound on $\nu=0$. For the $\nu=1/4$ system, as long as the asymptotic value of $z$ satisfies the condition $\frac{3\pi}{8}{<}z_{t\rightarrow\infty}$, the maximal zero point of $\zeta$, i.e. $a^0_\zeta$ will lie outside that of $dH/da$, i.e. $a^0_{H'}$.}
\label{figdHdaZeropoint}
\end{figure}

In the special case of $M_1=M_2$ or $\nu=1/4$
\beq{}
\frac{dH}{da}|^{\nu=1/4}_{\omega^2\approx GM/a^3}=
\frac{G M^2 (2{-}3GM/a)}{2 a^2 (4{-}3GM/a)^{3/2}}
\eeq
In the general case of $\nu{\leqslant}1/4$, the function form of $\frac{dH}{da}$ will be more complicated. But its basic feature of containing a zero point somewhere inside the region $\frac{3}{2}\leqslant\frac{GM}{a}\leqslant6$ is very robust, referring to FIG.\ref{figdHdaZeropoint} for illustration. Denoting this value of $a$ as
\beq{}
a^{0}_{H'}\equiv a:\frac{dH[a,\nu]}{da}=0
\eeq 
By equation \eqref{dadtGeneralExpression}, the zero value of $dH/da$ will manifest as a divergent point on $\frac{da}{dt}$. However such a divergence will not bring us disconnection on the variation of $a(t)$. So it can be cured by hand in the numeric integration of eq\eqref{dadtGeneralExpression}. For  latter uses, we define another two extra special values of $a$ through
\beq{}
a^\infty_H\equiv a: H[a,\nu]=\infty
\eeq
\beq{}
a^0_\zeta\equiv \max\{a:\zeta(t)\equiv\frac{\sin4GMz/a}{4GMz/a}=0\}
\eeq
where $a^\infty_H$ is the point on which $H$ itself diverge and $a^0_\zeta$ is the maximal zero point of $\zeta$.
In the special case of $\nu=\frac{1}{4}$, according to equation \eqref{HbinaryMassB}, we can easily see that $a^\infty_H=\frac{3GM}{4}$.
On the evolution line of $a(t)$, whether the divergence implied by $\frac{dH}{da}=0$ will be implemented depends on the relative relation between $a^0_{H'}$ and $a^0_{\zeta}$, see FIG.\ref{figdHdaZeropoint} for references. 

\begin{figure}[ht]
\includegraphics[totalheight=32mm]{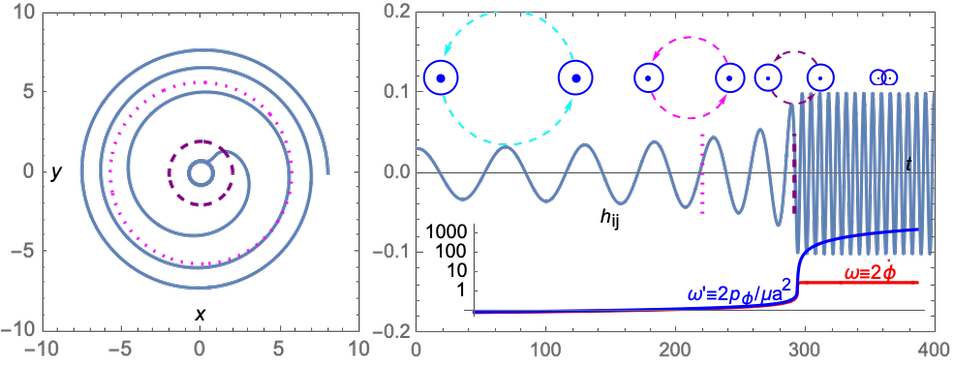}
\includegraphics[totalheight=32mm]{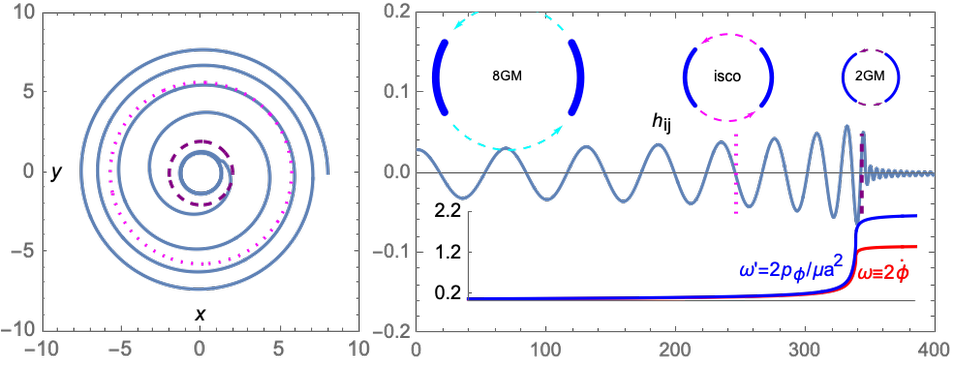}
\caption{The relative motion orbit and gravitational wave form of binary systems consisting of equal mass members following from XOB method. The participating black holes in the upper row have point like singular inner-structure thus are described by the standard Schwarzschild metric exactly. Their shape deformation parameter $z{\equiv}0$. The participants in the lower row have extended inner-structure but fixed length of $4GM$. So the shape deformation parameter $z{\equiv}1$. This inner-structure will experience bending but no elongation as the merger progresses. The dotted magenta line marks the ISCO radius defined by the EOB method. The dashed purple line is the mathematic horizon radius $r_h=2GM_{tot}$.}
\label{figGWshape0p0X}
\end{figure}

Since $\zeta{=}\frac{\sin 4GMz/a}{4GMz/a}$, the concrete value of $a^0_\zeta$ depends on the form of $z(t)$ inevitably. The standard Schwarzschild black hole is defined by the exact horizon and singularity. By the language of newonian mechanics, its matter distribution is singular and experiences no stretching or bending. By our definitions of banana shape deformation, such black holes have $z\equiv0$ and the corresponding radiation activity factor $\zeta\equiv1$. In the case the two members have equal mass, equation \eqref{dadtGeneralExpression} allows us to trace their merger evolution down to $a(t)\xrightarrow{t{\rightarrow}\infty}a^\infty_H=\frac{3}{4}GM$ and get 
\beq{}
\omega\equiv\dot{\phi}\xrightarrow{t{\rightarrow}\infty}\big[\frac{GM}{(3GM/4)^3}\big]^\frac{1}{2}=\big(\frac{4}{3}\big)^\frac{3}{2}(GM)^{-1}
\label{qnfrequency0p}
\eeq
\beq{}
H\xrightarrow{t{\rightarrow}\infty}-\frac{2M/3}{\sqrt{0}}\rightarrow-\infty
\label{hamiltonian0p}
\eeq
The former equals to half of the real part of the late time quasi-mormal mode's frequency $\omega^\mathrm{re}_{022}$. The latter has the explanation of the final black hole's total mass or energy. Its divergence and negativity is a reflection of the irrationality of the point particle type inner-structure picture of black holes. The upper part of FIG.\ref{figGWshape0p0X} displays the full three stage gravitational wave forms correspondingly. To make the wave forms readable, we reduced the late time frequency and amplitude of the gravitational wave by a factor of $0.3$. The decay feature of late time gravitational wave forms is non-visible in the time interval displayed. The logarithmic divergence feature of $\omega'\equiv p_\phi/\mu a^3$ is due to the point particle type inner structure of the black holes involved. So this figure tells us that, if black holes have point like singular structure, then we cannot get quasi-normal mode type late time feature on the gravitational wave form of their binary merger event.

Now let us consider the extended inner structure and non-singular matter distribution of black holes. As first step, we let the black hole have fixed length which equals to the diameter of the corresponding Schwarzschild horizon so that $4GMz{=}4GM$, or $z{\equiv}1$ constantly. Although the length is fixed, we let the extended black hole be bendable as the inspiral and merger progresses, so that the radiation activity factor varies with time in the most simple but still non-trivial way $\zeta=\frac{\sin4GM/a}{4GM/a}$, see equation \eqref{zetaDefinition} for the derivation this expression. When this radiation activity factor comes in, the asymptotic value of $a(t)$ is no longer the singular point $a^\infty_H$ of the hamiltonian, but $a^0_\zeta$. In the case the two participating black holes have equal mass, $a^\infty_H=\frac{3GM}{4}$, $a^0_{\zeta}=\frac{4GM}{\pi}$. As results
\beq{}
\omega\equiv\dot{\phi}\xrightarrow{t{\rightarrow}\infty}\big[\frac{GM}{(4GM/\pi)^3}\big]^\frac{1}{2}=\big(\frac{\pi}{4}\big)^\frac{3}{2}(GM)^{-1}
\eeq
\beq{}
H\xrightarrow{t{\rightarrow}\infty}\frac{M/2(4-\frac{\pi}{2})}{\sqrt{4-\frac{3\pi}{4}}}\rightarrow0.9473M
\label{hamiltonian0p}
\eeq
The former still corresponds to half of the real part of the late time quasi-normal mode's frequency $\omega^\mathrm{re}_{022}$. The latter equals to the final black hole's total mass or energy. Combining with the full process gravitational wave form displayed in the lower part of FIG.\ref{figGWshape0p0X}, we easily see that, relative to those following from the exact horizon wrapped point particle type inner-structure picture, the gravitational wave form following from inner-structure picture characterised by extended inner-mass distribution with fixed length but banana shape bendable and wrapped only by approximate or asymptotic horizons are much more reasonable, due to its late time quasi-normal mode type oscillation.

In the practical BHB merger process, the degree of the participating black holes' banana shape deformation varies with time.   Just as we pointed out in section \ref{secBanana}, when such banana shape deformation happens, the participating black holes will be elongated along the inspiral direction. So during the whole merger process,  we expect their length varies from zero to some final length value $2{\pi}a_f$. Substituting the value of this final length and the corresponding $\omega_f=\big(\frac{GM}{a_f^3}\big)^{1/2}$ into the hamiltonian \eqref{singlebodyHamiltonian}, we should get an energy less than the total mass of the initial black holes
 \beq{}
 H[a_{\!f},\nu,\big(\frac{GM}{a_{\!f}^3}\big)^\frac{1}{2}]<M
 \label{eqAfLowLimit}
 \eeq
This is because the gravitational wave radiation dissipates energy of the system inevitably. From this inequality, we can solve out the lower limit of $a^\mathrm{low}_\mathrm{f,lim}$ and a corresponding frequency $\omega_f$, which equals to half of the real part of the merged black hole's quasi-normal mode's frequency
\beq{}
\omega^\mathrm{re}_{022}\equiv2\dot{\phi}<2\big[\frac{GM}{a^\mathrm{low3}_\mathrm{f,lim}}\big]^\frac{1}{2}\equiv\omega^\mathrm{re022}_\mathrm{uplim}
\eeq
On the other hand, from the hamiltonian expression \eqref{HbinaryMassB} we note that given $\nu$ and $M$, $H$ is a lower bounded function of $a$, i.e. $H^{\nu,M}_\mathrm{min}{<}H^{\nu,M}(a)$. So we cannot set the mass of the final merged black hole to arbitrarily small value. Setting $H[a,\nu,(\frac{GM}{a^3})^\frac{1}{2}]$ to this minimal value $H^{\nu,M}_\mathrm{min}$, we will get an upper limit for the value value $a^\mathrm{upp}_\mathrm{f,lim}$ and a corresponding lower limit on the value of $\omega^\mathrm{re}_{022}$
\beq{}
\omega^\mathrm{re022}_\mathrm{lolim}\equiv\big(\frac{GM}{a^\mathrm{upp3}_\mathrm{f,lim}}\big)^\frac{1}{2}<
2\dot{\phi}\equiv\omega^\mathrm{re}_{022}
\eeq 
However, since we have not any physical principle to prohibit a black hole be banana shape deformed and elongated longer than $2\pi a^\mathrm{upp}_\mathrm{f,lim}$, so that the corresponding lower limit $\omega^\mathrm{re022}_\mathrm{lolim}$ can only be considered a qualitatively estimation for the value of $\omega^\mathrm{re}_{022}$.

\begin{figure}[ht]
\includegraphics[totalheight=45mm]{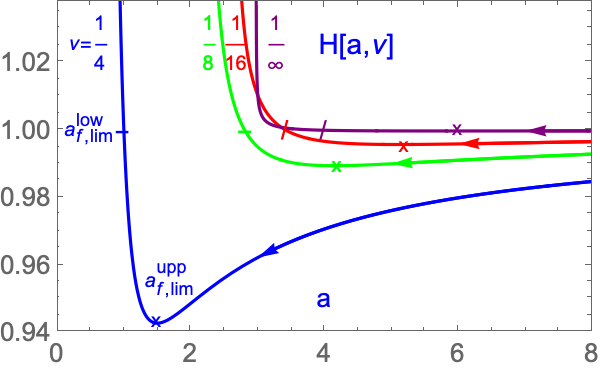}
\includegraphics[totalheight=45mm]{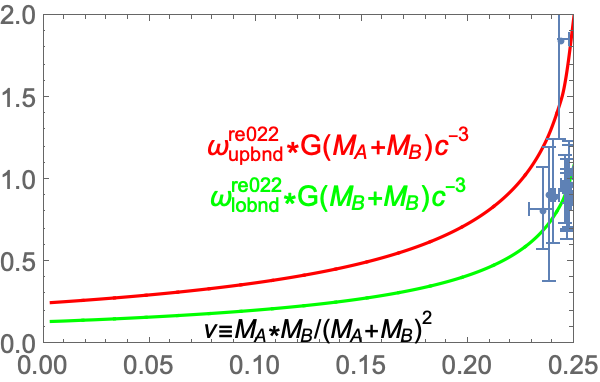}
\caption{The upper panel is the hamiltonian $H[a,\nu]$'s dependence on $a$ for four typical values of $\nu$. During the merger process of two black holes, the relative obit radius $a$ decreases continuously. Setting $a=a^\mathrm{low}_\mathrm{f,lim}$ will leads to an upper bound on the real part of quasi normal frequency $\omega^\mathrm{re}_{022}$; setting $a=a^\mathrm{upp}_\mathrm{f,lim}$ will give a lower bound on $\omega^\mathrm{re}_{022}$}
\label{figOmAfnllimit}
\end{figure}

The upper part of FIG.\ref{figOmAfnllimit} displays $H[a,\nu]$ and $\omega^\mathrm{re022}_\mathrm{uplim}$'s dependence on $a$. The lower part of the figure displays $\omega^\mathrm{re022}_\mathrm{lolim}$'s dependence on the symmetric mass ratio $\nu$. Also displayed on the downer part of the figure is the currently available observational data of \cite{GWTC1,GWTC2,GWTC3}. From the figure, we can easily see that, our definite prediction for $\omega^\mathrm{re022}_\mathrm{uplim}$ is perfectly satisfied by the observational data except one point. While $\omega^\mathrm{re022}_\mathrm{lolim}$ gives rather precise estimation for the value of $\omega^\mathrm{re}_{022}$. By this estimation, when $\nu=\frac{1}{4}$, the value of $a^\mathrm{upp}_{f,lim}=\frac{3}{2}GM$. This means that the two participating black holes are elongated to  $\pi a^\mathrm{upp}_\mathrm{f,lim}=\frac{3\pi}{2}GM$ as the merger finishes. When $\nu=0$, the value of $a^\mathrm{upp}_{f,lim}=6GM$. That is, the final black hole in this case should be considered a rotating ring of perimeter $6\pi GM$.

In principle, by tracing up the matter distributions' evolution under the radiation back reaction force's driving, it is possible to determine the function form of $z(t)$ theoretically. However, such a work is very difficult to accomplish in practice. So we will illustrate the observational feature of the black hole deformability's difference by comparing the gravitational wave forms following from two shape deformation modes of
\beq{}
z_1(t)=\frac{\pi a_f}{4GM}{\cdot}\frac{a_f}{a(t)},
z_2(t)=\frac{\pi a_f}{4GM}{\cdot}\big[\frac{a_f}{a(t)}\big]^2
\eeq
The corresponding radiation activity factors read
\beq{}
\zeta_1=\frac{\sin[\pi a_f^2a^{-2}]}{\pi a_f^2a^{-2}}
,
\zeta_2=\frac{\sin[\pi a_f^3a^{-3}]}{\pi a_f^3a^{-3}}
\eeq
The black holes characterised by $z_1(t)$ experience the banana shape deformation more intensively at the early stages of the merger process, while those characterised by $z_2(t)$ experience the shape deformation more intensively at the late stages of the merger process.  

\begin{figure}[h]
\includegraphics[totalheight=32mm]{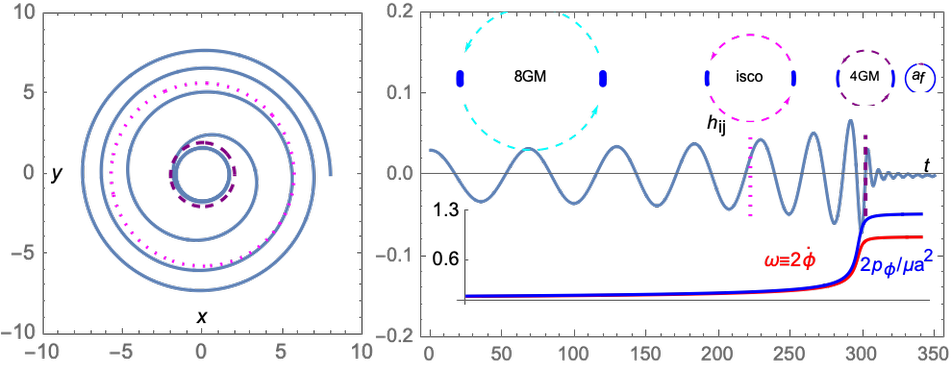}
\includegraphics[totalheight=32mm]{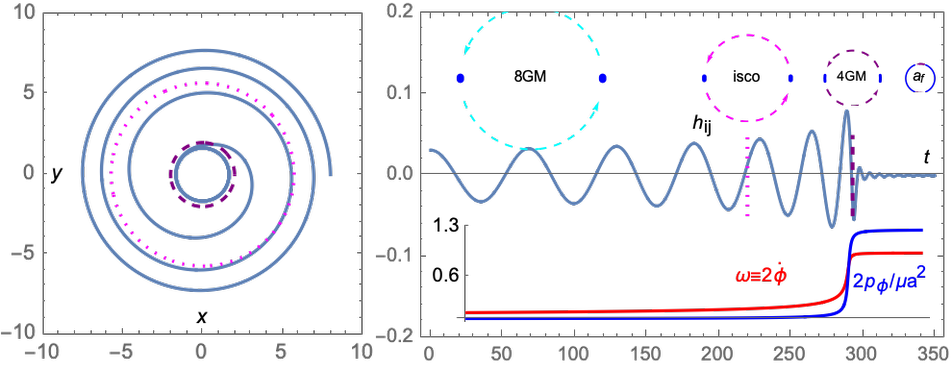}
\caption{The relative motion orbit and gravitational wave form of BHB systems consisting of equal mass members with inner structure and experience banana shape deformation. The banana shape deformation parameter in the upper black holes varies with time at the way $z{\propto}a_fa(t)^{-1}$, the lower varies at the way $z{\propto}a_f^2a(t)^{-2}$. The the gravitational wave amplitude in the lower figure not only grows more suddenly, but also decays more suddenly around the $a\approx2GM$ region. The dotted magenta line marks the ISCO radius defined by the conventional EOB method. The dashed red line is the horizon radius of the total mass black hole $r_h=2GM_{tot}$.}
\label{figGWshapeQNM1122}
\end{figure}

FIG.\ref{figGWshapeQNM1122} displays the relative motion orbit and the corresponding gravitational wave forms of these two types of black holes. From the figure, we can see that the $z{\sim}a(t)^{-1}$ black holes experience banana shape deformation more mildly and more intensively during the early stages of the inspiral motion, so their relative motion orbit radius evolves to the asymptotic value $a^0_f$ in longer time durations. Numerically this is $t^1_{isco}=222$, $t^1_h=302$ while $t^2_{isco}=220$, $t^2_h=293$. We will call this type of black holes softer than those characterised by $z{\sim}a(t)^{-2}$.  The most general form of the banana shape deformation function can be written as
\beq{}
z(t)=k_0+k_1{\cdot}\frac{a_f}{a(t)}+k_2{\cdot}\frac{a^2_f}{a(t)^2}+\cdots
\eeq
Except $a_f$, all other parameters $k_i$ in this function are dimensionless. For these dimensionless parameters, we conjecture that they are independent of the black hole sizes. By carefully measuring of the late time gravitational wave form following from various BHB merger process, we have chances to determine these parameters observationally. So wether the black holes in the nature have inner structures and how hard such inner structures are are a question answerable experimentally.

\section{Conclusion}
\label{secConclusion}

In one sentence, our conclusion can be stated that black holes in the nature have microscopic states understandable in standard general relativity and inner-structures testable from the gravitational wave forms produced in the BHB merger events, which is calculable analytically with our exact one body method. Expansions of this sentence are as follows.

We provided two exact time dependent solution families to the Einstein equation sourced by dust fluids and showed how such solution families form basis of black hole microscopic state both from the viewpoint of outside fixed position observers and from the viewpoint of inside co-moving observers. By the canonic quantisation scheme, we proved that the degeneration degree of such solution families has right exponential-area law feature as required by the Bekenstein-Hawking entropy formula except some polynomial corrections. Singularity theorem judges impossible to understand black hole microscopic state in general relativity only because, in applying such a theorem, the horizon and singularity's observer dependence is neglected under the name of physical reality's general coordinate invariant.

A time dependent collapsing star solution indeed contains a future singularity and an asymptotically realisable horizon.  But such a singularity and horizon would not become realities in any finite future of the outside fixed position observer by their physic time definition. They will be realised in the finite future only to the freely falling co-moving observers which use the proper time of the collapsing matter themselves as time definition. In another word, the horizon and singularity are physical realities only to the super investigators who adopt the global viewpoint or have the ability to see the future of infinitely far away future of the outside fixed position observers. To physical probe or observers such as the participants of BHB merger events or freely falling observers co-moving with the matter contents of a collapsing star, the horizon and singularity of their being observed objects are either non-implementable in any finite future or although reachable in finite future but non-detectable experimentally. The belief that black holes in the real world carry exact horizon and singularity makes two mistakes of (A) taking the horizon and singularity of the global viewpoint holders as the horizon and singularity of the local viewpoint holders  and (B) taking the physical law's general coordinate invariance as the physical reality's observer independence, which is not the case at all. 

Since either participant involved in a BHB's merger event can only see their merger partner's regular inner structures characterised by the approximately linear mass function and an asymptotic horizon, both of them will experience banana shape deformation under the inhomogeneous back reaction arising from gravitational wave radiations. We introduced the concept of radiation activity $\zeta$ to quantify this deformation caused suppression of the system ability to emit gravitational waves and derived out that $\zeta=\frac{\sin4GMz/a}{4GMz/a}$, where $z$ is the elongation times of the banana shape deformed black holes relative to their shape deformation before diameter. $\zeta$ will be multiplied on the quadrupole of the binary system when the quadrupole formula is used to calculate the gravitational wave radiations of the system.

We then proposed an exact one body-lisation method to transform the general relativity two-body problem into a single body one. This method gives up the static spherical symmetric background of EOB and the conservative part of the system's action calls no post newtonian approximation as input. It puts the two merger participants in a three partitied spacetime which rotates synchronously with the merger participants. The metric of the spacetime region that accommodate the merger participants can be determined non-perturbatively by requiring that the two participant's inspiral motion is synchronous and central fixed. This method avoids the intrinsic inconsistence of the conventional EOB and provides descriptions for the relative motion of BHB systems throughout the full inspiral, merger and ring-down stages. Taking the quadrupole radiation as the source of dissipation, and neglecting the inner structure of black holes by looking them as  point particles wrapped by exact horizons, we calculated the relative motion orbits and the gravitational wave forms following from the BHB merger process and get results which are highly consistent with those from  conventional EOB but behave more rationally at the very late time stages.

When considering the extended inner structure of black holes, even the most simple fixed length banana shape deformation model, the gravitational wave forms following from our XOB method are almost the same as those following from the combination of EOB+NR+BHPT methods and measured in the real observations. We derived out an exact upper bound on the real part of the final black holes' quasi-normal modes $\omega^{\mathrm{re}022}_\mathrm{upbnd}$ irrespective of the black hole inner structure details as a concrete prediction of XOB. We also derived out a not so strong lower bound  $\omega^\mathrm{re022}_\mathrm{lobnd}$ for the late time quasi normal modes which can be used as an estimation for the typical value of of $\omega^\mathrm{re}_{022}$. Putting the currently available data on the theoretical prediction figure we see very good agreement between the two. We compared gravitational wave forms following from BHBs whose members are characterised by different shape deformation susceptibilities and point out that the differences between them may probably be measurable in the current and future observations. So the question on what inner structures the physic black hole has is answerable observationally.

Two possible directions for future works would be, to find observational evidences for the black hole inner structures we proposed here through other channel such as black hole images and gravitational wave echos,  and to revise and apply our XOB method to BHBs or neutron star binaries whose members are spinning and carry nonzero angular momentum.

\newpage
\section*{Acknowledgements}
We thank very much to Asta Heinesen, Maarten van de Meent, Gregorio Carullo, Rodrigo Panosso, Daniel D'Orazio, Vitor Cardoso and Zhengwen Liu for their comments on the manuscript of this work and everyday communications. This work is done in Niels Bohr International Acadmia. We thank very much to the Chinese Scholarship Council for their sponsor of oversea visiting researches and the warm hosts provided by Vitor Cardoso and Julie de Molade. This work is supported by CSC202006545026, NSFC grant no. {11875082}, 
, the Villum Investigator program of VILLUM Foundation (grant no. VIL37766) and the DNRF Chair program (grant no. DNRF162) by the Danish National Research Foundation.

\end{document}